\def\deg{^{\circ}} 
\def\kms{\ifmmode{\hbox{km~s}^{-1}}\else{km~s$^{-1}$}\fi}
\def\la{\mathrel{\hbox{\rlap{\hbox{\lower4pt\hbox{$\sim$}}}\raise1pt\hbox{$<$}}}}
\def\ga{\mathrel{\hbox{\rlap{\hbox{\lower4pt\hbox{$\sim$}}}\raise1pt\hbox{$>$}}}}

\documentstyle[12pt,psfig]{article}

\begin{document}

\begin{center}

{\Large {\bf The Ursa Major Cluster of Galaxies. II. \\
\medskip
Bimodality of the Distribution of Central Surface Brightnesses}}

\vspace{1.5cm}

R. Brent Tully$^1$ and Marc A.W. Verheijen$^2$ \\
\bigskip
$^1$ Institute for Astronomy, University of Hawaii, 2680 Woodlawn Drive,
Honolulu, HI 96822 \\ e-mail: tully@ifa.hawaii.edu \\
\smallskip
$^2$ Kapteyn Astronomical Institute, Postbus 800, NL-9700 AV Groningen,
The Netherlands \\ e-mail:verheyen@astro.rug.nl \\

\end{center}

\vspace{1.0cm}

\noindent{ABSTRACT---} The Ursa Major Cluster appears to be unevolved
and made up of HI-rich spiral galaxies like one finds in the field. 
$B,R,I,K^{\prime}$ photometry has been obtained for 79 galaxies,
including 62 in a complete sample with $M_B^{b,i} < -16.5^m$ (with a
distance to the cluster of 15.5~Mpc).  The $K^{\prime}$ information is
particularly important for the present discussion because it is not
seriously affected by obscuration.  There is reasonably convincing
evidence that the distribution of exponential disk central surface
brightnesses is {\it bimodal}.  There is roughly an order of magnitude
difference in the mean luminosity densities of high and low surface
brightness disks.  Disks {\it avoid} the domain between the high and low
surface brightness zones.  The few intermediate surface brightness
examples in the sample all have significant neighbors within a projected
distance of 80 kpc.  The high surface brightness galaxies exhibit a
range $-21^m < M_B^{b,i} < -17^m$ while the low surface brightness
galaxies are found with $-19^m < M_B^{b,i}$ down to the completion
limit.  High and low surface brightness galaxies in the overlap regime
$-19^m < M_B^{b,i} < -17^m$ that are indistinguishable in
luminosity--line width plots have very distinct locations in surface
brightness--scale length plots.  The existence of separate high and low
surface brightness families suggests that there are discrete radial
configurations that are stable.  Galaxies are driven into one of these
regimes.  The high surface brightness state has a lower luminosity
cutoff.  It is likely that the high surface brightness galaxies are
dominated by dissipational matter at their centers while the low surface
brightness galaxies are dark matter dominated.  The high surface
brightness family subdivides into those with, and without, substantial
bulges.  In either case, these galaxies have essentially the same
exponential disk central surface brightnesses.  Evidently, there are
{\it two thresholds} probably controlled by angular momentum content or
transfer.  Passing from high to low specific angular momentum, there is
first the transition from low surface brightness to high surface
brightness regimes, then the transition from exponential disk to disk
plus bulge regimes.


\newpage

\section{Freeman's Law}

Evidence will be presented that the luminosity densities of the disks
of galaxies are bimodally distributed.  By inference, the mass
densities would also be bimodally distributed.  If true, then the
distinction between high and low surface brightness systems is a
fundamental phenomenon that any theory of galaxy formation should be
required to explain. 

The surface brightness of galaxy disks falls off exponentially with
radius (de Vaucouleurs 1959).  Freeman (1970) considered many of the
most familiar nearby spiral and S0 galaxies and pointed out that if
fits to the main body of disks are extrapolated to the centers 
then the central surface brightnesses have
values in the $B$ 
band of $\mu_0^B \simeq 21.65$ magnitudes per square arcsec with a very
small dispersion.  Today many exceptions are known to Freeman's law,
inevitably on the low surface brightness side.  It has been argued
that the constancy of central surface brightness values was an
artifact either of obscuration (Jura 1980), or of bulge-disk
decomposition methods (Kormendy 1977), or of selection effects that
narrow the range of observed surface brightnesses (Disney 1976; Allen
\& Shu 1979).  
Van der Kruit (1987) made a strong case for a peak in agreement with
Freeman's discovery, although he appreciated that there are low
surface brightness galaxies that are discordant.
The current dominant point-of-view is that there 
is a rather flat distribution of central surface brightnesses from
 $\mu_0^B \sim 20$ down to
the faintest accessible levels (McGaugh, Bothun, \& Schombert 1995; 
McGaugh 1996; de Jong 1996$a$).
The details of this distribution have not been well
determined because of selection effects.  

Wide field photographic
surveys select for high surface brightness
galaxies within a preferred `visibility' window, in the vocabulary of
Disney \& Phillipps (1983).  Special efforts are necessary to find low surface
brightness objects (Impey, Bothun, \& Malin 1988; Irwin et al. 1990;
Davies et al. 1994; Schwartzenberg et al. 1995).  These efforts amply
demonstrate the ubiquitousness of low surface brightness galaxies but
often the distances to these objects are unknown and it
is not easy to achieve a normalization of the counts per surface
brightness bin per unit volume element.
Going back a bit, the first big compilation
of low surface brightness galaxies was by van den Bergh (1959, 1966).
Redshifts for these galaxies were obtained by Fisher \& Tully (1975)
which made it clear that many of the objects in this class are big and
intrinsically luminous; the brightest in that sample have $M_B \sim
-20$.  A big compendium of nearby low surface brightness
galaxies is provided by Fisher \& Tully (1981).  Quite a few samples
have been compiled since that time, for example, the new catalog by
Impey et al. (1996) which includes objects to $z \sim 0.1$.

This new study contains two observational elements that appear to be
important.  One is the nature of the sample: we have a statistically
significant data set which is {\it complete to an absolute magnitude
limit} drawn from an environment dominated by HI-rich, mostly non-interacting
disk galaxies.  The other element has to do with the passbands of the
photometric material: we combine $K^{\prime}$ imaging with $B,R,I$ 
optical imaging.

The raw data has been published in Paper~I of this series (Tully et
al. 1996).  There is a description of the Ursa
Major Cluster in that reference.  The region is unusual in a favorable
way.  Clusters 
are nice because if objects are at a common distance then an {\it
apparent} threshold corresponds to an {\it absolute} threshold.
However, it is suspected that cluster environments affect the
properties of the constituent galaxies, for example, to produce the
correlation of galaxy types with local density (Dressler 1980).  Yet
it was argued in Paper~I that the Ursa Major Cluster may be so young
that it's members are representative of a `field' population.  The
velocity dispersion of the cluster is only 148~\kms, whence the
characteristic crossing time is 0.5~H$_0^{-1}$.  There is no
concentration toward a core, no large ellipticals and only a few
moderate S0's, and the spirals have normal gas properties.  Next to
the Virgo Cluster, the Ursa Major Cluster contains by far the largest
concentration of spiral galaxies of any bound region in the Local
Supercluster.  Hence, there is the possibility that all galaxies
in the volume of the cluster can be surveyed down to some absolute
limit and that these galaxies will be reasonably representative of
objects in the low density parts of the universe.

The $K^{\prime}$ ($2.2 \mu$m) imaging provides the second important resource
for this study.  A $256 \times 256$ HgCdTe detector was used in 
wide-field modes, providing fields-of-view up to 9~arcmin.  The photometry
in the infrared is not significantly affected by obscuration and it
measures the light from old populations, which is presumably more
closely tied to the total mass than the light of young populations.

It has long been appreciated that Freeman's law, if true, provides
strong constraints on galaxy formation scenarios.  However a galaxy
modeler could not be sure if (s)he was being asked to reproduce a
real effect or an observational artifact.  It is hoped that the
present discussion will help revive the focus on an important
characteristic of galaxies with the introduction of a couple of
curious new elements. 

\section{Surface Brightness -- Scale Length Diagrams}

A compilation of photometric data can be found in Paper~I for 79
galaxies projected 
within a $7.5\deg$ circle and with $700 < V_{helio} + 300 {\rm sin}
{\ell} {\rm cos} b < 1210$ \kms.  Of these, 62 galaxies define a {\it
complete sample} with $M_B^{b,i} < -16.5$.  CCD $B,R,I$ photometry is
available for all 79 galaxies.  There is imaging $K^{\prime}$
photometry available for 60 of the 62 galaxies in the complete sample
and for 10 of the 17 fainter galaxies.  Absolute
magnitudes are based on an assumed distance modulus of 30.95,
corresponding to a distance of 15.5 Mpc. Hence, $1^{\prime \prime} =
75$ pc.  This distance is compatible with a complex velocity field map
of the Local Supercluster with a global value of H$_0 = 85$
\kms~Mpc$^{-1}$ (Tully et al. 1997).

\subsection{Photometric parameters}

One-dimensional projected surface brightness profiles were derived by
averaging the surface brightness in concentric ellipses of constant
position angle and ellipticity chosen to match the outer isophotes.
The surface brightness profiles were fit by a straight line of the
form 
\begin{equation}
\mu^{\lambda}(r) = \mu_0^{\lambda} + 1.086 (r/r_d) ,
\end{equation}
corresponding to an exponential profile of the form
\begin{equation}
L^{\lambda}(r) = L^{\lambda}_0 e^{-r/r_d}.
\end{equation}
If a bulge was detected, the inner region of the profile was excluded
from the fit and $\mu_0$ was determined by extrapolation.  Hence,
the fall-off of surface brightness with radius $\mu(r)$ in a given
passband $\lambda$ is described by two parameters: the surface
brightness at the center of the galaxy $\mu_0$ and the exponential
scale length $r_d$.  Typically, 4 to 6 scale lengths are observed above
the outer limiting isophot.  
It is known that giant elliptical
galaxies are poorly fit by an exponential form (de Vaucouleurs
1959) but there are no giant ellipticals in the Ursa Major Cluster.
There are some moderate luminosity early-type systems that are
classified as S0.  It is possible to provide at least a crude
exponential fit to the main bodies of these galaxies, although
inevitably there are substantial bulge components at the centers.
It is known that dwarf galaxies, whether irregular or
elliptical, can be reasonably approximated by exponential profiles
(Binggeli, Sandage, \& Tarenghi 1984).  Hence, it has been possible to
measure the parameters $\mu_0$ and $r_d$ in all available bands for all
the galaxies in the sample.

Plots of the positions of galaxies in the domain of
these two parameters are informative.  We begin in Figure~1 with the
directly observed 
parameters; ie, with no corrections for inclination effects.  The
solid symbols denote members of the complete sample and the crosses are
associated with fainter galaxies.  The diagonal lines are loci of
constant magnitudes since for an exponential disk
\begin{equation}
m_T^{\lambda} = \mu_0^{\lambda} -2.5 {\rm log} 2\pi(b/a)
-5 {\rm log} r_d .
\end{equation}
where the observed minor-to-major axial ratio of a galaxy is $b/a$.
The lines are drawn to coincide with the completion
boundary, defined in $B$ and roughly transformed to the other bands.
The curved lines in the $B$ panel are in the spirit of the visibility
limits discussed by Disney \& Phillipps (1983).  Following from
Paper~I, the magnitude above the limiting isophote is
\begin{equation}
m_{lim} = m_T - 2.5 {\rm log}[1-(1+{(\mu_{lim}-\mu_0) \over 1.086})
e^{-{(\mu_{lim}-\mu_0) \over 1.086}}] .
\end{equation}
Here, $(\mu_{lim}-\mu_0) / 1.086$ is the number of scale lengths above
the limiting isophote $\mu_{lim}$.  If we
substitute for $m_T$ and assume the face-on case $b/a=1$,
\begin{equation}
{\rm log}r_d = 0.2 [ \mu_0 - 2.5 {\rm log}2\pi - m_{lim} -
2.5{\rm log}[1-(1+{(\mu_{lim}-\mu_0) \over
1.086})e^{-{(\mu_{lim}-\mu_0) \over 1.086}}].
\end{equation}
We consider that we have completion brighter than $m_{lim}=14.5^m$. 
Our fundamental reference is 
the Uppsala General Catalogue of Galaxies (Nilson 1973).  According
to Cornell et al. (1987), this catalog measures diameters to 
$25.4^m \pm 0.7$ at $B$ and
according to van der Kruit (1987) it measures diameters
to $26.0^m \pm 0.7$ at $B$.  In the following, we assume there is
completion to $\mu_{lim}^B = 25.5^m$.  With these
assumptions for $m_{lim}$ and $\mu_{lim}$, Eq.~(5) provides the
relation between $r_d$ and $\mu_0$ plotted as the lower limiting curve
in Fig.~1.  The upper left limiting curve is imposed by the
requirement that galaxies have diameters larger than 1~arcmin at the
limiting isophote.  In this case,
\begin{equation}
r_d (\mu_{lim}-\mu_0)/1.086 > 30^{\prime\prime}
\end{equation}
that is, the scale length times the number of scale lengths above the
limiting isophote must exceed a radius of $30^{\prime\prime}$.

The galaxies of the complete sample seem already to fall into two
zones in each of the color panels of Fig.~1.  As one progresses from
$B$, through $R$ and $I$, to $K^{\prime}$, the two zones become more
separated.  The higher surface brightness galaxies tend to
be redder so the differences between high and low surface brightness
objects are accentuated with observations toward the infrared.

\subsection{Inclination corrections}

Inclination effects can be confusing at optical bands because surface
brightness pathlength and obscuration variables play off against each
other.   Happily, at $K^{\prime}$ obscuration is negligible and it can
be anticipated that projection effects on surface brightness are
simply described by geometric considerations.  It is expected that the
central surface brightnesses of galaxies viewed face-on can be
described by
\begin{equation}
\mu_0^{\lambda,i} = \mu_0^{\lambda} - 2.5 C^{\lambda} {\rm log} (b/a)
\end{equation}
Here, the coefficient $C$ ranges from 0 for an opaque system to 1 for a
transparent system.  The superscript $i$ means that an inclination
correction has been applied.  It can be anticipated that
$C^{K^{\prime}} \sim 1$ at 
$K^{\prime}$ and $C^{\lambda}$ is progressively smaller as one goes toward
shorter wavelengths.  In transparent systems the geometric effect of longer 
line-of-sight pathlengths in 
edge-on cases augments surface brightnesses, but if the systems are not 
transparent the geometric augmentation
is off-set by the increased obscuration in edge-on cases.

In fact, if one distinguishes between edge-on and face-on galaxies in
Fig.~1 (triangles and circles, respectively), there is an immediately
evident separation between the two 
inclination groups on the $K^{\prime}$ plot that weakens at $I$ and $R$ and
is almost washed out at $B$.  It is exactly this effect that is
anticipated by the formulation of Eq.~(7).  Estimates of $C^{\lambda}$
can be derived by looking for the best agreement between edge-on
and face-on galaxies in the various passbands.  Given the apparent
separation of galaxies into two surface brightness zones, we split the
sample at $\mu_0^{K^{\prime}} = 17.5$ and, moreover, consider only the 62
galaxies of the complete sample.  For the high and low surface
brightness sub-samples separately, we then varied $C^{\lambda}$
to find minima in the dispersion of central surface brightnesses.
Figure~2 illustrates the variations in rms dispersion with the choice
of the parameter $C$, for the high surface brightness sub-sample in panel
$a$ and for the low surface brightness sub-sample in panel $b$.  The
variation of $C^{\lambda}$ behaves as expected for the 39 galaxies in
the high surface brightness 
sub-sample (38 at $K^{\prime}$).  There is a minimum dispersion at
$K^{\prime}$ with  
$C^{K^{\prime}}=1$ corresponding to the transparent model.
Progressively toward shorter wavelengths, dispersion minima occur at
$C^I=0.61$, $C^R=0.52$, and $C^B=0.23$.  At minimum, the rms
dispersions are $\sim 0.53^m$ at $R$ and $I$ and $\sim 0.58^m$ at $B$
and $K^{\prime}$.  For comparison, Valentijn (1990) found $C^B \sim
0.2$ for S$b$-S$c$ types and Peletier \& Willner (1992) found $C$ near the
transparent regime at $H$-band.

If the same test is applied to the 23 galaxies of the low surface
brightness sub-sample (22 at $K^{\prime}$), the results are more
uncertain but consistent 
with the proposition that these systems are transparent.  It is seen
in Fig.~2$b$ that the rms dispersion is minimized with $C^{\lambda}$
in the range 0.64-0.78, with no systematic dependence on $\lambda$.
Since $C^{\lambda}$ does not increase with increasing $\lambda$ we
conclude that 
obscuration is not a factor for this sub-sample.  The deviation from
$C=1$ is taken to be a statistical aberration.  Almost certainly,
$C^{K^{\prime}}$ should equal unity and the measured minimum is most deviant
from unity in this case.  We accept that $C^{\lambda}=1$ at all
passbands for the low surface brightness sub-sample.  Dispersions
about the mean surface brightnesses are $0.5^m$ to $0.7^m$ but these
values may be affected by incompletion on the low surface brightness
side.  Surface brightness means and dispersions are recorded for the
various subsamples and passbands in Table~1.

Figure~3 is the equivalent of Fig.~1 but now with the inclination
adjustments of Eq.~(7) applied.  In this new figure, the high surface
brightness (HSB) 
galaxies are distinguished as the filled symbols and the low surface
brightness (LSB) galaxies are identified by open symbols.  The
separation between the two groups is formally made at
$\mu_0^{K^{\prime},i}=18.5^m$.  In the $K^{\prime}$ panel, large symbols
identify galaxies with large bulges (concentration index from Paper~I 
$C_{82} > 5$).
It is seen that the HSB and LSB
domains are each restricted in surface brightness and extended in
scale length.  There are histograms of the surface brightness
distributions shown in Figure~4.  The one HSB system, as defined at 
$K^{\prime}$, that overlaps with the LSB sample at $B,R,I$ is the
anomalous galaxy NGC~3718 that will be discussed later.

These plots include the adjustments
for inclination effects, so care is needed in one regard.  In the cases
of $B,R,I$ the coefficients $C^{\lambda}$ are different for the HSB
and LSB regimes (see Table~1).  The effect of the inclination
corrections are to make the adjusted surface brightnesses of edge-on
galaxies {\it fainter}: $\mu_0^{\lambda,i} \geq \mu_0^{\lambda}$.  The
larger the value of $C^{\lambda}$ the larger the shift.  In the case
of the $B,R,I$ passbands, since $C_{LSB}^{\lambda} > C_{HSB}^{\lambda}$
there is a {\it separation} of the high and low surface brightness
groups introduced by the inclination adjustments.
On the one hand, there is a reasonable basis for making the separate
inclination corrections to HSB and LSB systems.  On the other hand, it
makes it dangerous to argue that the gap between HSB and LSB systems
is real when part of the separation is introduced by these
corrections.  Hence, the $K^{\prime}$ material takes on a particular
importance.  Our tests indicate that both the HSB and LSB galaxies are
in the transparent regime, so the two groups receive the same
inclination treatment.  Moreover, the HSB systems turn out to be
redder than the LSB systems so the separation between the two kinds of
galaxies is most easily distinguished in the infrared.

The color and reddening variations provide a reconciliation with the
claims by Peletier \& Willner (1992) that the range of observed
surface brightness at $B$ is small because of dust absorption and is
more considerable at $H$ where disks are almost transparent.  If no
corrections for inclination are made then the scatter is smallest at
$B$ because absorption and projection effects off-set each other.
Once suitable corrections are made then the scatter is comparable in
each band from $B$ to $K^{\prime}$ for the separate HSB and LSB families.
However the HSB and LSB families move apart as one progresses from the blue
to the infrared since HSB types are redder than LSB types.  Hence, if
the separate families are not 
distinguished then the dispersion in surface brightnesses seems to
increase as one progresses to the infrared.

Could the bimodality be an artifact of our fitting of the exponential
disks since we have not attempted bulge--disk decompositions?  The large 
bulge systems are flagged in Fig.~3 and the bimodality is seen to remain
in those without bulges.  It is argued by de Jong (1996$b$) and Courteau,
de Jong, \& Broeils (1996) that the exponential representation of bulges
is at least as justified as an $r^{1/4}$ representation, whence de Jong 
shows our ``marking the disk'' fits give an unbiased disk characterization
relative to a bulge/disk decomposition, with an uncertainty of $\sim 0.2^m$.
A greater potential error arises in our case because the $K^{\prime}$
photometry is cut off at smaller radii by sky noise.  If there is a 
substantial bulge, the disk fitting range is restricted and there is a
bias toward too steep a slope through inclusion of some of the bulge in the 
disk.  Hence, the measure of 
$\mu_0^{K^{\prime}}$ may be too bright.   The concern here is whether 
bulge galaxies could 
have been moved from the gap to the HSB domain erroneously.  The
$K^{\prime}$ luminosity profile fits of Paper~I have been reconsidered from 
a conservative perspective by asking how far $\mu_0^{K^{\prime}}$ values 
could be pushed toward the gap.  In fact, in several of the bulge systems 
it is warranted to take fainter $\mu_0^{K^{\prime}}$ and larger 
$r_d^{K^{\prime}}$.  Changes from Paper~I are recorded in Table~2.  A
couple of systems are moved into the surface brightness gap but the
changes are not significant.

\subsection{An environmental effect}

Having identified the separate HSB and LSB families we wondered whether 
there was any environment difference between the two, so we
looked at surface brightness properties as a function of proximity to 
nearest neighbor.  Figure~5 shows the amazing result.  In the top panel 
for each bandpass,
the inclination adjusted central disk surface brightness is plotted against
the projected distance to the nearest significant neighbor. To be 
`significant', we require that the neighbor have at least 10\% of the 
luminosity of the galaxy under consideration.  For this discussion, we draw
attention to the $K^{\prime}$ panels where the situation is clearest.
Remarkably, all of 
the intermediate surface brightness systems have projected near neighbors.
{\it For the $\sim 2/3$ of the sample that do not have a significant close 
companion the separation into HSB and LSB classes is compelling.}

The middle panels for Fig.~5 repeats the histograms of Fig.~4 
but only includes the 38 of 62 galaxies (36 of 60 at $K^{\prime}$) in the 
complete sample with
nearest significant neighbor more distant than 80~kpc in projection.
Histogram means and dispersions are recorded in Table~1.  The solid
curve in the $K^{\prime}$ panel illustrates a completeness
expectation.  The histogram would have this shape if there was a
uniform population of the $\mu_0^{K^{\prime},i} - r_d$ domain for
$0.8<{\rm log} r_d<1.6$ and $\mu_0^{K^{\prime},i}>17^m$.  The fall-off
from the peak is described by the transposition of the curve defined
by Eq.~(5) and illustrated in the $B$ panel of Fig.~1.
The bottom panels in Fig.~5 shows the same information for the 24 galaxies 
in the complete sample with a nearest significant neighbor closer than
80~kpc in projection.  

For the isolated galaxies, there is a gap at $K^{\prime}$ between HSB and 
LSB types of $1.5^m$, which contains 3 galaxies where roughly 20 might
be expected.
The dispersion about the separate peaks is $\sigma = 0.40$.  There is
an evident
difference with respect to the representative completeness expectation.
This remarkable figure demonstrates that the HSB-LSB bimodality is 
highly significant in galaxies that are relatively isolated today.
These galaxies have transit times $> 10^9$ years with another galaxy.

The sample is drawn from a cluster but one that appears to be dynamically 
young; so much so that in Paper~I it was argued that the galaxies may be 
representative of a field population.  The relatively isolated galaxies 
which display the bimodality most clearly are probably most similar to 
galaxies in the field.  By contrast, the much more scattered
distribution of surface brightnesses for the galaxies with projected
neighbors is strong evidence that interactions can substantially 
redistribute the luminous matter in disks.

We can summarize this section with the suggestion that the disks of
galaxies tend to be in either a high surface brightness state or a low
surface brightness state and avoid the intermediate ground.  At
$K^{\prime}$, the difference between the mean central surface
brightness of the disk components in each state is a full factor of
10.  The evidence for bimodality is particularly strong if only
relatively isolated galaxies are considered.  The high 
surface brightness family lie sufficiently far from the completion
limits on the $\mu_0 - r_d$ plots that our census of this family is
probably near to complete.  On the contrary, the low surface
brightness family is badly intersected by the completion limits and
our census of that family must be quite incomplete.

\section{Back to the Literature}

It has been appreciated for some time that there are departures from
Freeman's law, to the extent that the acronyms LSB and HSB have
become familiar coinage for low and high surface brightness systems.
However it had never been proposed that there was a {\it discrete
difference} between LSB and HSB galaxies.  Rather, it was supposed that
there was a continuum of surface brightness properties (cf, McGaugh 1996;
de Jong 1996$a$) and the designations LSB and HSB described 
objects on either side of an ill-defined dividing line (Davies et
al. 1988$a$; McGaugh \& Bothun 1994; de Blok, van der Hulst, \& Bothun
1995).  See McGaugh (1996) for a more elaborate catagorization.

In retrospect, astronomers have long been able to distinguish LSB from
HSB galaxies on a qualitative basis.  Figure~6 illustrates the
correlation between surface brightness class and morphological type
designations.  With overwhelming coincidence, galaxies typed S0--S$c$
are classed HSB and galaxies typed S$cd$--I$m$ are classed LSB.  The few
exceptions tend to be rather anomalous and hard to define
morphologically.  The HSB/LSB separation at $\mu_0^{K^{\prime},i}=18.5$
takes the S0 galaxy NGC~4117 to the LSB class but this limit is 
subjective and could be revised.
Evidently, the density of the disk has a distinct
signature in the appearance of a galaxy.  For one thing, the HSB
systems may saturate at the centers on photographic images while the
LSB systems do not.  There must also be manifestations in the
organization of spiral structure.

Our claim of bimodality is consistent with the study by van der Kruit
(1987).  His sample was somewhat smaller, with distance effects, and
was based on $J$-band photographic material with quoted uncertainties
in $\mu_0$ of $\pm 0.3^m$.  He saw enough of a difference between
early and late types that he fit separate distributions to the two,
but he did not draw attention to these separations or make any
suggestion that the two types were distinct.  Van der Kruit made
inclination adjustments on the assumption that the disks are
transparent which closes the gap in $\mu_0$ between the HSB and LSB
families.  Consider the $B$-band panel of our Fig.~5.  If we assumed
HSB galaxies are transparent at $B$ then the HSB sub-sample would
shift to lower surface brightnesses (more positive values) by $0.6^m$ in
the mean.  The LSB sub-sample remains where it is because it was
already assumed that these galaxies are transparent.  Hence, the two
sub-samples would begin to merge.  The mean for the HSB part would be
$\sim 21.2^m$, not so different from van der Kruit's $\sim 21.5^m$ for
S0--S$c$II (translating from $J$ to $B$) and Freeman's (1970)
$21.65^m$ for a blend of mostly HSB and a few LSB objects.

De Jong (1996$a$) presents a similar plot to our Fig.~6 with his Fig.~3.
It can be seen, on the one hand, that our results are consistent with 
his and, on the other hand, that his sample of mostly earlier types
would not convincingly reveal bimodality.

If we are seriously proposing a distinct separation of disk types into
two families, then it is an appropriate moment to try to understand
the relationship between these families and the kinds of galaxies that
have been revealed by other surveys.  In terms of
luminosity--surface brightness--scale length properties it might be
argued that there are 
as many as {\it six} families of galaxy types.

{\it Type 1: giant boxy ellipticals.}  Galaxies of this class do not
have significant disks and are not well described by the exponential
luminosity-radius description.  If exponential curves were force-fit
in such cases, presumably these systems would be given $\mu_0$ values
at least as bright or brighter than HSB disks and comparable
scale lengths.  There are no such galaxies in our sample.

{\it Type 2: high surface brightness disks.}  Normal spiral and S0
galaxies, the majority of our complete sample, are of the HSB type.
It can be debated if `disky' ellipticals belong in this group or
with type 1.

{\it Type 3: low surface brightness disks.}  Galaxies typed S$cd$ to
Irregular are of this LSB class, which constitutes a third of our
complete sample and a half of our overall sample.  The examples we
know about are inevitably HI-rich.  Our study does not explore the
full domain of this class at faint $\mu_0$ and low $r_d$.

{\it Type 4: dwarf spheroidals.}  Galaxies of this type are known in
the Local Group and related objects are found in abundance in such
clusters as Virgo (Binggeli et al. 1984; Impey et al. 1988) and Fornax
(Ferguson \& Sandage 1988; Davies et al. 1988$a$; Irwin et al. 1990).
The vaste majority of 
these systems are HI-poor.  The location of this class in the 
$\mu_0 - r_d$ parameter space is shown in Figure~7 with the superposition
of the Fornax Cluster samples of Davies et al. (1988$a$) and Irwin et
al. (1990).  The
domain of these dwarfs is 
essentially entirely below our completion limit.   Galaxies of types 3
and 4 must overlap in surface brightness and scale length properties
and the relationship between the two groups remains to be clarified.
It has been argued (Wirth \& Gallagher 1984; Kormendy 1985) that dwarf 
spheroidals are a distinct family from large ellipticals.

{\it Type 5: compact dwarfs.}  Blue and red compact dwarfs are known
to exist (Zwicky 1964).  The best example of a blue compact in our
sample  is PGC~37045 = 1148+48 = Markarian 1460.  UGC~6805 may be a
reasonable example of a red compact.  These two objects are fainter than
our completion limit.  There is still a very poor inventory of these
kinds of galaxies.  Red compacts may be related to giant ellipticals
and blue compacts may be related to dwarf irregulars.

{\it Type 6: large low surface brightness galaxies.}  Malin~1 (Bothun
et al. 1987) is an extreme example of what seems yet to be a rare
class of galaxies.  Other examples have been reported by Davies,
Phillipps, \& Disney (1988$b$) and Sprayberry et al. (1995) and
representatives of these
objects have been located in Fig.~7.  Galaxies of this class can have
extreme properties because of low values of $\mu_0$ and large values
of $r_d$.
The objects known to date are distinguishable from galaxies of type~3
because they are relatively {\it red} and have {\it prominent
bulges}.  Both types 3 and 6 contain HI.  It is possible that we have
one galaxy of this type in our 
sample, albeit not so extreme.  NGC~3718 is quite anomalous in
comparison with the rest of our objects.  It has by far the largest
exponential scale length in all passbands and has a relatively low disk
central surface brightness.  It is red and has a big bulge which
causes it to be classified S$a$ but the galaxy could be considered too
pathological to be fit into the Hubble sequence.  Schwarz (1985) has
shown that this otherwise relatively isolated galaxy could be
interacting with NGC~3729.  It has not been suggested in other cases that
the large low surface brightness class are involved in interactions.

In summary of this section, galaxies with distinctive properties
inhabit distinctive parts of the $\mu_0 - r_d$ diagram.  In
particular, galaxies we think of as `normal' disk systems are
reasonably {\it isolated} from other types in this parameter space.
These normal galaxies can 
be quantitatively specified by their $\mu_0$, $r_d$ properties.  We
should ask how this segregation of properties has occurred.

\section{A Core Surface Brightness -- Luminosity Relation}

This section stands a bit apart.  The disk component can get lost at
the centers of some galaxies that have large bulges.  In Paper~I,
we tabulated for each galaxy the surface brightness within an ellipse
with a major axis 
radius of $4^{\prime\prime}$ and a minor axis radius in proportion to
the inclination of the galaxy.  At the distance of Ursa Major,
$4^{\prime\prime} = 300$ pc.  This parameter that we call $\mu_4$ is a
metric surface brightness, a measure of the density of light at the
centers of all our galaxies.  It is the cummulation of the light of
disks and of any bulge.  Active nuclei, and possibly bars, will
contribute to $\mu_4$.  At $K^{\prime}$, the parameter should not be
strongly affected by obscuration.

Tight correlations are seen in Figure~8 which shows all the available
data at $K^{\prime}$ band.  In the top panel, there is the relationship
between the $\mu_4^i$ parameter and the disk central surface brightness,
$\mu_0^i$, where the superscript indicates a correction for
inclination has been made.  This inclination correction is applied to
the disk component only; ie, the flux from the bulge contribution to
$\mu_4$ is not modified with inclination.  The two parameters,
$\mu_0^i$ and $\mu_4^i$, scatter about the 45 degree equality line  
if there is no bulge component but $\mu_4^i < \mu_0^i$ if there is a bulge
component in addition to the disk component.  Only types earlier or
equal to S$ab$ among the HSB sub-sample have significant bulge
components.  In panels $b$ and $c$, the $K^{\prime}$ magnitude is plotted
against the $\mu_4^i$ core surface brightness parameter and tight
correlations are found.
The same data are shown in the two panels but in $b$
the types S$ab$ and earlier are emphasized with big symbols and in $c$
the types S$b$ and later are emphasized with big symbols.  The
correlations are particularly tight with these separations by type.

These plots are a diversion from our main theme but it is worthwhile
to remember that, while the disk central surface brightnesses of HSB
galaxies may have a small scatter, the disk-plus-bulge (plus possible
active nucleus) central surface
brightnesses display a wide range.  In our small sample, there are
separate strong correlations between $\mu_4^i$ and $M_{K^{\prime}}$ for
types S0--S$ab$ and S$b$--I$m$.

\section{Luminosity Functions}

A convenient description of the luminosity function of galaxies is
provided by the Schechter (1976) formulation.  However, there is
growing evidence that this two-parameter curve is too simple.  Samples
that are fit only to $M_B \la -16^m$ are adequately described with the
Schechter parameter $\alpha \simeq -1.0$ (Davis \& Huchra 1982; Tully
1988; Loveday et al. 1992; Marzke et al. 1994) which is the case if
there are equal numbers in equal logarithmic bins at faint
luminosities.  However, there have been claims that the luminosity
function turns up, possibly dramatically, at the faint end (Sandage et
al. 1985; Driver et al. 1994; Marzke et al. 1994).  In other words,
the luminosity function for all galaxies combined may have a concave
shape that the Schechter formulation will not accommodate.

It is interesting to see the separate contributions to the total
luminosity function according to the HSB and LSB catagories that have
been identified.  Figure~9 shows the luminosity functions of the
separate HSB and LSB components and the sum of all types.  Overall,
the luminosity function is rather flat to the completion limit of
$M_B=-16.5^m$.  The separate components have very different forms.
The HSB component {\it cuts off above the completion limit} while the
LSB component is rising sharply at the completion limit.

This result is not surprising given the strong correlation between the
surface brightness classes and morphological types, as shown in
Fig.~6.  Sandage, Binggeli, \& Tammann (1985) have demonstrated the
differences in luminosity functions between morphological types.  See
also Binggeli, Sandage, \& Tammann (1988) for a review, and Marzke et
al. (1994).  The Binggeli et al. review recalls the historical debate
over the bell-shaped function found by Hubble (1936) versus the
faint-end exponential shape advocated by Zwicky (1942).  It has been
appreciated that Hubble was drawing upon a sample dominated by high
surface brightness objects while Zwicky was impressed that low surface
brightness, faint galaxies did exist but were strongly selected
against.  There are some basic points of agreement.  Binggeli et
al. (1988) find bell-shaped luminosity functions for types E--S$c$, as
we do for the HSB family, and as Hubble found for samples dominated by
HSB galaxies.  Binggeli et al. and Marzke et al. (1994) find
luminosity functions to be steeply increasing at the faint end for
irregular or dwarf spheroidal systems, as we find for the LSB family and
as Zwicky anticipated.

The HSB--LSB distinction does put a new twist on the debate.  It was
not clear before why one should be impressed by the decomposition of the
luminosity function by type carried out by Binggeli et al.  Let us
restrict our 
considerations to either the HI-rich disk systems or the HI-poor
ellipsoidal systems separately.  If there is a continuum of properties
along one of these branches then the type decomposition may just be
providing an alternative description of the Hubble sequence rather
than telling us something fundamental about galaxy formation.  For
example, suppose the sequence from S$a$ to I$m$ is basically a
continuous mass sequence.  More massive galaxies are more organized,
have bigger bulge components, get an earlier type classification, and
are more luminous.  The least massive, less luminous galaxies are too
small to maintain spiral structure and have a late type
classification.  Naturally then, there would be differences in the
luminosity functions of the different types and earlier types will
have cut-offs at the faint end.  The luminosity function differences
between types might  just be telling us about thresholds for the
maintenance of spiral structure or the formation of bulges: issues to
do with resonances or disk instabilities, perhaps.  The ensemble
luminosity function might be viewed as providing a more global
constraint on the mass spectrum and galaxy formation.

The HSB--LSB dichotomy provides a better understanding of the origins
of the historic debate and enhances the interest in the separation of
luminosity functions by type.  If there really were a continuum of
surface brightness properties among disk systems then the extreme bias
of early and even modern surveys in favor of E--S$c$ types at the expense
of S$d$--I$m$ types is hard to understand, the `visibility' arguments
of Disney \& Phillipps (1983) not withstanding.  The situation makes
more sense if there is a distinct gap in the detectability of the
early and late types, as follows from what we are finding.  Moreover,
the gap has to be explained.  The step from S$c$ to S$d$ may involve a
discontinuity in formation processes.  For example, what if there is a
step in mass-to-light ratio between the S$c$ and S$d$ types, a possibility
discussed in the next section.  Then the ensemble luminosity function
would not be a simple reflection of the more fundamental mass
function.  A 
kink in the ensemble luminosity function between the domains of HSB
and LSB dominance might be a signature of a break between different
ways that disks form.  

\section{Luminosity -- Line Width Relations}

Although the possibility of surface brightness bimodality has come as
a surprise, part of the original motivation for our study was to
understand why galaxies can lie together on a luminosity--HI profile
line width plot (Tully \& Fisher 1977) yet be quite removed from each
other on a surface brightness--scale length plot.  The
luminosity--line width relations are seen in Figure~10 for the fraction
of our sample that satisfies our type, inclination, and HI profile
quality criteria (34 galaxies).  The HSB galaxies are distinguished by
filled symbols; the LSB galaxies, by open symbols.

In all the passbands, it can be seen that there is a $2^m-3^m$ domain
of overlap between HSB and LSB galaxies and that there is no
significant difference of each type from the mean correlations.  The
HSB systems are redder than the LSB systems so the HSB types do
progressively brighten relative to the LSB's as one steps toward the
infrared.  The galaxies in the overlap region are given box symbols so
they can be tracked in the next plot, Figure~11.  This figure is a
repetition of Fig.~3 except now the galaxies given box symbols in
Fig.~10 are identified with identical symbols here.  These figures
illustrate the point made in the previous paragraph.  Galaxies that
are mixed together in luminosity--line width diagrams 
separate to the distinct HSB and LSB zones on the $\mu_0-r_d$ diagrams.

Three specific HSB--LSB pairs are illustrated in Figure~12.  Each of
these pairs lies close together on the luminosity-line width diagrams.
See the objects labeled 1,2,3 in Figs. 10 and 11.  The $I$-band
inclination-adjusted surface brightnesses are shown in Fig.~12 and the
distinct differences between the HSB and LSB families are evident.

The HSB-LSB overlap in Fig.~10 is part of the mystery of the tightness of
luminosity--line width relations.  The coincidence between LSB and HSB
systems has been noted by Sprayberry et al. (1995) and Zwaan et
al. (1995).  Zwaan et al. make the following point if galaxies are
built homologously.  Compare systems with the same luminosity $L$ and
maximum rotation velocity $V_{max}^{obs}$ but different $\mu_0,r_d$.  
Then the indicative mass,
$M \propto (V_{max}^{obs})^2 r_d$, is greater for the LSB galaxy with 
large $r_d$
than the HSB galaxy with small $r_d$, though the luminosities are the
same.  That is, $(M/L)_{LSB} > (M/L)_{HSB}$.  This argument is in line
with other evidence that late-type galaxies have
larger mass-to-light ratios than early-types (Carignan \& Freeman
1988; Persic \& Salucci 1988).

If the velocity field is responding to just the matter distributed in
an exponential disk then the velocity maximum induced by this disk,
$V_{max}^{disk}$, will occur 
at $2.1 r_d$ (Freeman 1970).  With this consideration, we can
generalize that there are
three possible cases: (i) if the potential within a few $r_d$
is dominated by mass in an exponential disk then the
observed $V_{max}^{obs}$ should occur at $r \sim 2.1 r_d$,
(ii) if the potential is dominated by the halo
$V_{max}^{obs}$ will probably occur at $r > 2.1 r_d$, 
since the halo is expected to be more extended than the light,
and modeling will probably give $V_{max}^{disk} \ll V_{max}^{obs}$, 
and (iii) if in addition to the self-gravitating exponential disk
there is a central bulge then $V_{max}^{obs}$ should be pulled inward to
occur at $r < 2.1 r_d$. 

We can look at the information about velocity fields in the literature.
The HSB--LSB pair NGC~2403--UGC~128 was studied by de Blok \& McGaugh 
(1996) because they share the same $V_{max}^{obs}$ and $L$.  The
HSB object NGC~2403 has an exponential disk scale length about a third
that of the LSB object UGC~128.  The luminosity is sufficiently
concentrated in the HSB case that the stellar mass that can be associated 
with the light is enough to produce the observed velocities.  Such is not
the situation with the LSB case.  The
resolution in the case of UGC~128 is not fully satisfactory (beam
$\sim 2r_d$).
Verheijen (1997) has Westerbork HI synthesis maps of all the
galaxies, both HSB and LSB, in the overlap region of $-17^m > M_B >
-19^m$ of the Ursa Major sample.  Figure~13 provides a preview of a small
bit of the HI information available to us.  Velocity--radius contour
maps are shown for the triplet of galaxies tracked in Figs. 10 and
11.  

The three galaxies used in this figure were chosen as representatives
of the three generalized cases outlined two paragraphs above.  We will
try to make the case that NGC~3949 is an example of a self-gravitating
exponential disk system without a substantial bulge.  This galaxy is in 
the HSB family.  It is offered as an example of case (i) above and will
be referred to as the "exponential HSB" class.  UGC~6973 has an HSB
disk and, additionally, has a significant central excess of light.  It 
is offered as an example of case~(iii), the "exponential HSB plus bulge"
class.  NGC~3917 is drawn from the LSB family.  It will be shown that
the disks of such galaxies do not make an important dynamical
contribution.  This object is an example of case~(ii), the "LSB" 
class.

The bottom panels of Fig.~13 demonstrate decompositions of the rotation
curves according to the contributions associated with the stellar component,
assuming $M/L_{K^{\prime}} = 0.4 M_{\odot}/L_{\odot}$, the interstellar
gas component (negligible in all these examples), and what is left over,
hence attributed to a dark halo.  The choice $M/L_{K^{\prime}} = 0.4
M_{\odot}/L_{\odot}$ was made to give a `maximum disk' fit (van Albada
\& Sancisi 1986) of the velocities associated with the luminous matter
to the observed inner rotation velocities for UGC~6973 and NGC~3949, the
two HSB examples.  There was no attempt to make a bulge-disk
decomposition.  For the LSB NGC~3917, the same choice of $M/L$ fails by
a wide margin to explain the observed rotation.  The disk contribution
could be raised with an increased $M/L_{K^{\prime}}$ but not
sufficiently for it to become dominant.  If anything, it might be expected that
$M/L$ values are lower for LSB systems compared with HSB types, not
higher.  The idea that there is a dynamical difference between HSB and
LSB types is revisited in the next section.  The photometric gap appears
to have a dynamical correspondence. 

\section{Discussion}

A discreteness between `normal' and `dwarf' galaxies was suggested by
Dekel \& Silk (1986) based on a theoretical idea, though the idea was
motivated by observations.  They argued for a potential well {\it
threshold}: small galaxies loose much of their gas when an early burst
of star formation leads to supernova driven winds that exceed the
escape velocity, while large galaxies retain the gas.

The concept could explain the distinction between LSB and HSB
conditions and the apparent augmentation in $M/L$ with LSB types.  It
may be a problem for the theory that there are examples of the LSB
class within our small sample that have rotation velocities as high as
$V_{max}^{obs} \sim 150$ \kms, since Dekel and Silk predict there is
substantial mass lose only if virial velocities are
$\la 100$ \kms.
Perhaps that stretch can
be accommodated.  However, we have a more fundamental concern.

It is not evident how the Dekel--Silk model fits with the observation
that HSB and LSB systems of a given maximum rotation value have the 
same luminosity.
Their `dwarfs' loose a large fraction of the gas
that could create stars and luminosity.  Alternatively, these
observations could be satisfied if two conditions are met: (i) {\it
the maximum velocity of a galaxy is set by the dark halo}, and (ii)
{\it there is a fixed initial fraction of the protogalaxy in
matter that can form stars and this matter is largely conserved.}
This latter condition is quite at odds with the Dekel--Silk
proposition, although remember that we need only be concerned with the
domain $-19^m < M_B < -17^m$ where HSB and LSB types overlap.  The
Dekel--Silk mechanism could kick in at a fainter magnitude.

If our two rules are met then the observations of Figs. 10 and 11 can
be explained since a predictable number of stars are formed in a given
halo, though the concentration may vary.  We need a {\it different}
mechanism from that proposed by Dekel and Silk, one that creates the
discrete HSB and LSB classes while {\it preserving} the dissipational
mass content.

Mestel (1963) pointed out that self-gravitating disks would find their
ways to specific radial configurations since there are radial forces
in disks that are not experienced if there is spherical symmetry.
Mestel anticipated that one stable radial gradient results in a flat
rotation curve. 
Gunn (1982) has shown that a flat rotation curve due to a dynamically
important disk embedded in an isothermal halo implies an exponential mass
distribution for the disk.  Ryden \& Gunn (1987) get flat
rotation curves with galaxy formation in a cold dark matter scenario
but the halos are dominant.  Ryden (1988) could find an inner regime
where the dissipational material is dynamically dominant in cases with
large initial fluctuation amplitudes which become large galaxies.
These models do not entertain possible transfer of angular momentum
between successive collapsing shells during formation.

Our proposition is that the HSB and LSB modes correspond to two
alternative radially stable configurations.  To understand the two
possibilities, let us consider the large and small extremes, then the
intermediate cases.  The small extreme may be the simplest.  Suppose
that the dissipational component has sufficient angular momentum that
it reaches rotational equilibrium at densities that still leave the halo
dominant.  If small galaxies that have avoided merging are the progeny
of small amplitude ($\sim 1 \sigma$) initial fluctuations then it is
reasonable that they formed late, hence with relatively high
specific angular momentum (Efstathiou \& Jones 1979)

One of Mestel's (1963) stable configurations involves self-gravitating
diffuse disks but there is evidence from disk--halo decompositions
that the diffuse LSB disks are {\it not} self-gravitating (Fig.~13, 
NGC~3917, and de Blok
\& McGaugh 1996).  The dark matter halos contribute substantially at
all radii.  If this
quasi-spherical system collapsed conserving angular
momentum in radial shells then it ends up in the stable configuration
described by Ryden and Gunn.

The luminous galaxies find their way to become disk dominated at their
centers.  Somehow we end up with the disk--halo `conspiracy' of flat
rotation (van 
Albada \& Sancisi 1986).  There have been a lot of experiments with
N-body simulations in the framework of the cold dark matter model and
there is general agreement with the observed rotation properties of
galaxies (cf, Blumenthal et al. 1986; Navarro, Frenk, \& White 1996).
There are systematics 
that make the `conspiracy' less than perfect, with some initial decline
with radius in concentrated, luminous systems (Casertano \& van Gorkom
1991).  The central densities of the disks remarkably conform to
Freeman's law.  Galaxies know the law.  If there is too much
dissipational material in the cores to abide the law then a bulge is
formed.  The `conspiracy' co-ops this third component.

Our qualitative interpretation of events is that the galaxies have
settled into a mandated stable radial configuration,
in the spirit of the Mestel argument but with the revision required
due to the lurking dark halo.  The formation process presumably
involved a
considerable transfer of angular momentum outward to the halo.  Such a
process would seem to be a natural consequence of hierarchical
merging where blobs come together in non-axisymmetric ways, and is
seen in N-body simulations (Barnes \& Efstathiou 1987) and collision
simulations with gas (Barnes \& Hernquist 1991).  The differences
in the surface brightness distributions for galaxies with, and without,
close neighbors, seen in Fig.~5, provides strong
evidence that interactions can reorganize the luminous matter in 
galaxies.  Even without such jostling,
the Ryden (1988) models show that more massive systems (with parts
that form earlier with less angular momentum) have more significant
dissipationally dominant cores.  The dissipational collapse of disks
was discussed by Fall \& Efstathiou (1980) and we can mention
the contribution by Shaya \& Tully (1984).

Two stable radial configurations have been identified.  In massive
systems with a significant component of low angular momentum material,
the dissipational matter will collapse to form a dynamically
important disk that rotates with a velocity tied to the
requirements of the dark halo.  In dwarf systems the dissipational
material does not collapse enough to dominate the dark halo even near
the center.  Now consider the intermediate regime.  The evidence from
the Ursa Major Cluster sample is that {\it galaxies opt for one of the
two aforementioned stable configurations rather than find some state
in between.} 
There is a two to three magnitude domain between giants and dwarfs 
where the
galaxies have a choice.  It is seen from Fig.~9 that, proceeding from 
bright galaxies to faint, the number of HSBs drop and the number of
LSBs picks up.
The evolutionary path followed by a specific
intermediate-size galaxy must depend either on the amount of
angular momentum it acquired as a protocloud or on the degree of
trauma of it's birth and during it's lifetime.  Systems that retain
lots of angular momentum 
hang up as LSB galaxies.  Systems that never had much angular
momentum or, probably more relevant, those that transfer angular
momentum outward collapse to the HSB state.

The focus has been on the distinction between LSB and HSB families.
There is the second transition between systems that are exponential
disks to their cores and systems with central bulges.  Both varieties
have disks that obey Freeman's law.  Densities do not want to exceed
the mandated threshold 
in a rotating disk.  If there is low angular momentum material that
would cause an excess then it finds its way into a bulge.

What we have been presenting at first blush seems to contradict the
`universal rotation curve' idea of Persic, Salucci, \& Stel (1996).
How can our three discrete states be reconciled with their continuum
of rotation properties as a function of total luminosity?  The
apparent contradiction at the state transition
between bulge/no--bulge can be disregarded because Persic et al. make
the disclaimer that their `universal rotation curve' may not apply at
small radii in cases of bulges.  As for the state transition between
HSB and LSB, it depends on how one looks at the data.  In Fig.~13 it
is seen that HSB and LSB systems have very distinct {\it metric}
radial distributions because the LSB types are much more extended.
However the rotation curves are only different in a subtle way if the
radial 
scale is {\it normalized} by either exponential scale length or
optical radius as Persic et al. do.  The exponential HSB NGC~3949 
reachs $V_{max}^{obs}$ at 
$\sim 2.1 r_d$, located by the little arrow in the lower panel of
Fig. 13.   The LSB NGC~3917 reachs
$V_{max}^{obs}$ somewhat farther out than this photometric scale 
length located by the arrow.  It will take a large sample of high
quality velocity fields to confirm if the rotation curves are
continuing to rise at $\sim 2.1 r_d$ in LSB types while they tend 
to have peaked
by this radius in HSB systems of the corresponding luminosity.
Since the HSB and LSB types overlap in luminosity,
it can be understood how the very small differences in normalized
curves get averaged and result in a smoothly continuous `universal
rotation curve'.  

It is important to recall the point made by Zwaan et
al. (1995), however.  The masses and $M/L$ measures are functions of
the {\it metric} radii.  Hence, though the scale length--normalized
rotation curves of an HSB--LSB pair of a given luminosity might be
almost indistinguishable, still $(M/L)_{LSB} > (M/L)_{HSB}$ within the
optical domains.  In the
case of the HSB galaxies, it was seen in Fig.~13 that model velocities 
$V_{max}^{disk}$ approaching $V_{max}^{obs}$ can  
be associated with the light and $M/L$ values characteristic of star
ensembles.  By contrast, for LSB systems the velocities
$V_{max}^{disk}$ associated with stars  
and gas fall far short of $V_{max}^{obs}$ (see also de Blok \& McGaugh 
1996, 1997).  
Qualitatively, $V_{max}^{disk}$ is given by the relationship:
\begin{equation}
{(V_{max}^{disk})^2 \over L} \sim {(M/L) \over r_d} .
\end{equation}
In an HSB system, $r_d$ for a given $L$ is small and the required
$M/L$ to give $V_{max}^{disk} \sim V_{max}^{obs}$ 
is reasonable.  In an LSB system with the same $L$ and
$V_{max}^{obs}$, there is
a larger $r_d$ and the required $M/L$ to give $V_{max}^{disk} \sim
V_{max}^{obs}$ is unreasonably large for a stellar system.  If
anything, the blue LSBs can be expected to have {\it lower} $M/L$
values than redder HSBs.

Note
that Persic et al. (1996) are arguing as we do that low luminosity galaxies
have much more important dark halo contributions within the optical
domain than is the case with high luminosity galaxies.  The difference
is that they see the transformation with luminosity as a continuum
while we think there is a discrete transition involved.  Our
point-of-view receives some support from information in the
literature.  As in Fig.~13, disk--bulge decompositions of 
well-established rotation
curves lead to models where {\it either} the disk component
substantially dominates the halo interior to $\sim 2 r_d$ {\it or} the
disk component is at best comparable to the halo at the center.
We are unaware of a well established case where the disk is only
modestly dominant at the center.

\section{Linkage Between Photometric and Kinematic Properties}

A direct linkage can be drawn between the surface brightness bimodality
and a dynamical bimodality.  The luminosity--line width fits given by
the lines in Fig.~10 translate to relationships between rotation velocities, 
surface brightnesses, and scale lengths given Eq.~(3) and our assumed
distance to the sample of 15.5~Mpc.  At $K^{\prime}$ band:
\begin{equation}
{\rm log} W_R^i/2 = 3.355 - 0.112 \mu_0^{K^{\prime},i} 
+ 0.561 {\rm log} r_d^{K^{\prime}}.
\end{equation}
Here, $W_R^i/2$ approximates $V_{max}^{obs}$ and, to be precise, this 
relation is based on the double regression to the data in Fig.~10 
rather than the single regression with errors in $W_R^i$ that we use for 
distance measurements.
At the same time, the peak rotation velocity that arises out of an exponential
disk is given by:
\begin{equation}
V_{max}^{disk} = 8.60 \times 10^4 \sqrt{10^{-0.4 \mu_0^{K^{\prime},i}} 
r_d^{K^{\prime}} M/L_{K^{\prime}}}
\end{equation}
or in the logarithmic form:
\begin{equation}
{\rm log} V_{max}^{disk} = 4.934 - 0.2 \mu_0^{K^{\prime},i} 
+ 0.5 {\rm log} r_d^{K^{\prime}} + 0.5 {\rm log} M/L_{K^{\prime}}
\end{equation}
where $V_{max}^{disk}$ in \kms\ is the peak rotation velocity at $2.1 r_d$
for an exponential disk, in the units of Figs. 1, 3, and 11 at our assumed 
distance.  The coefficient of Eq.~(10) is $3.14 \times 10^5$ if the units
of $r_d$ is kpc.  Combining these equations, 
\begin{equation}
{\rm log} (V_{max}^{disk}/0.5 W_R^i) = 1.579 - 0.0878 \mu_0^{K^{\prime},i} 
- 0.0610 {\rm log} r_d^{K^{\prime}} + 0.5 {\rm log} M/L_{K^{\prime}}.
\end{equation}
Suppose a fixed value of $M/L_{K^{\prime}}$ is assumed.  The maximum disk fits
illustrated in Fig.~13 suggest  $M/L_{K^{\prime}} = 0.4 M_{\odot}/L_{\odot}$.
Then a choice of $V_{max}^{disk}/0.5 W_R^i$ translates to an almost
horizontal line in Fig.~11.  The solid line illustrates 
$V_{max}^{disk}/0.5 W_R^i = 2/3$ and the dashed line illustrates
$V_{max}^{disk}/0.5 W_R^i = 1/3$.  The lines would be exactly horizontal
if the coefficient for the term in ${\rm log} r_d$ in Eq.~(9) were 0.5, 
which would arise if $L_{K^{\prime}} \sim (W_R^i)^4$.  Instead the
double regression of the $K^{\prime}$ panel of Fig.~10 gives 
$L_{K^{\prime}} \sim (W_R^i)^{3.6}$.  These lines make graphic the point 
first discussed by Aaronson, Huchra, \& Mould (1979): the relation
$L \sim (W_R^i)^4$ is explained if $\mu_0$ and $M/L$ are constants for disk
galaxies and $V_{max}^{obs} \sim V_{max}^{disk}$.

Eq.~(12) tells us that the bimodality in $\mu_0$ implies a bimodality in
either $V_{max}^{disk}/0.5 W_R^i$ or $M/L$.  A bimodality in $M/L$ is
unlikely since it would require that, at all bands from $B$ to $K^{\prime}$,
$M/L$ is greater (substantially!) for LSB galaxies though these systems are 
known to have
proportionately younger populations than HSB galaxies.  It is much more
probable that the bimodality is in $V_{max}^{disk}/0.5 W_R^i$.

Accepting the simplifying assumption that $M/L={\rm constant}$ 
(specifically, $0.4 M_{\odot}/L_{\odot}$ at $K^{\prime}$), then any 
location in the 
$\mu_0, r_d$ plane corresponds to a determinate vale of 
$V_{max}^{disk}/0.5 W_R^i$ from Eq.~(12).   Figure~14 is the histogram
of all such values for the isolated fraction of our sample (those with
no significant
neighbors within 80~kpc in projection).  Except for the weak dependence
on $r_d$ identified in Eq.~(12), Fig.~14 is essentially a rescaling
of the middle $K^{\prime}$ panel of Fig.~5.

The rotation curve decompositions of Fig.~13 help with the interpretation
of this rescaling of the bimodality phenomenon.  With the HSB cases,
$V_{max}^{disk}$ is a substantial fraction of $V_{max}^{obs}$.  Halo
components contribute but the disk components are dominant inside
$2 r_d$.  For example, with NGC~3949 the decomposition gives
$V_{max}^{disk}/V_{max}^{obs} = 0.68$.  By contrast, in the LSB case,
NGC~3917, $V_{max}^{disk}/V_{max}^{obs} = 0.39$.  It can be seen in
Figs.~10 and 11 that NGC~3917 is one of the most luminous LSBs and at
the high surface brightness limit of the LSB family.  Yet even in this
case, the dark halo potential dominates the disk at all radii.
Hence the bimodal distribution in Fig.~14 can be
given a dynamical interpretation.  The galaxies in the peak about
$V_{max}^{disk}/0.5 W_R^i = 0.73$ have self-gravitating disks with
rotation in response to the luminous matter at $r \la 2 r_d$.  The 
galaxies in the peak about $V_{max}^{disk}/0.5 W_R^i = 0.42$ have
dynamically insignificant disks in dominant dark halos.  These two
dynamical classes have direct correspondences with the photometric
HSB and LSB classes, respectively.

A qualitative story has been presented.  More detailed velocity field
data are required to go further, and of course more quantitative
modeling.  There is the usual disclaimer that we have assumed
Newtonian gravity is valid in this regime, for if not then the
discussion of this section is built on a bogus premise (Milgrom 1983;
Sanders 1984; 
Mannheim 1993).  Still, the bimodal surface brightness observations
have to be explained.

\section{Summary}

There is suggestive evidence that galaxies {\it avoid a region of
parameter space} between high central surface brightness and low
central surface brightness domains.  The statistics provided by the
present sample are insufficient to make the case compelling, however
it will be straightforward to check our claim with larger controlled
data sets.  Since we are predicting a minimum between two
distributions, any experimental criteria that includes both the HSB
and LSB regimes should properly sample the intermediate zone.  The
case for bimodality of surface brightnesses is strongest for galaxies 
without near ($<80$~kpc) neighbors.

We argue that {\it the luminosity function for the high surface
brightness family cuts off at the faint end.}  Again our results have
some uncertainty because the cut-off is close to our completion
boundary.  It will be easy to design an experiment that goes a
magnitude deeper and checks this claim.

There is the curiosity that {\it galaxies of a given luminosity,
whether of high surface brightness or of low surface brightness, have
little dispersion in HI profile width.}  Yet the light of HSB and LSB
galaxies is distributed so differently that it would be surprising if
the rotation curves of these classes were not correspondingly
different.  Preliminary evidence is presented that suggests that the
`exponential HSB' class have rotation curves that peak by 
$\sim 2.1 r_d$, as anticipated for dynamically important exponential
disks, while there is evidence that the `LSB' class have rotation
curves that continue to rise beyond this radius.  In the HSB types it 
appears that reasonable stellar $M/L$ values produce inner rotation
velocities that rise close to $V_{max}^{obs}$, while for LSB types 
velocities
associated with reasonable models of the disk components fall
considerably short of $V_{max}^{obs}$.

The dichotomy between HSB and LSB types implies to us that there are
two basic control conditions: (i) maximum rotation is mandated by the
dark halo, and (ii) there is a fixed fraction of precollapse
dissipational material that is preserved to give the observed
luminosity.  The latter point is at odds with the idea that the LSB
class is the result of gas loss during an early starburst phase, but
is the obvious explanation of the tightness of luminosity--line width
relations.  The HSB and LSB types of a given luminosity would live in
similar halos and have the same amount of dissipational material.  However
the LSB types are in rotational equilibrium in potentials dominated by
the dark halos at all radii, while the HSB types have either
transferred angular momentum away from much of their gas or were born
with low specific angular momentum and have secularly evolved to
stable states with dynamically important disks.  Though in our model
HSB and LSB types of a given luminosity have the same total
dissipational and dark halo masses, the HSB cases have lower {\it
measured} $M/L$ because the light is more confined within the dark
halo; ie, a smaller fraction of the total halo is within the domain of
the observed rotation curve.

While the emphasis has been on the distinction between HSB and LSB
families, attention has been drawn to another transition between
`exponential HSB' and `bulge HSB' families.  Freeman's law is obeyed
in both cases.  Evidently disks are forbidden from rising in central
density above this threshold.  Low angular momentum matter that would
broach this limit is shunted into a bulge.  The bulges have a clear
manifestation in the rotation curves.  Velocities approach maximum
values much closer to the centers than the radius $\sim 2.1 r_d$
characteristic of the peak for self-gravitating exponential disks.
There could be said to be the maintenance of the flat rotation curve
`conspiracy'.

It is pointless to go too far with speculation in the absence of much
detailed rotation curve information, but the potential importance of
these results should be given attention.  Only discrete radial
gradients of matter are stable for disk systems since, unlike in
spherical systems, a particle in orbit at a given radius feels the net
gravity of mass at larger radii.  Galaxies must arrange themselves
into one of these discrete states.  Evidently, galaxies of
intermediate mass have a choice.

\section{Acknowledgements}

Mike Pierce, Jia-Sheng Huang, and Richard Wainscoat participated in the
collection of data.
We thank Renzo Sancisi, Erwin de Blok, and the referee Stacy McGaugh for 
helpful comments.
This research has been supported by NATO Collaborative Research Grant
940271 and grants from the US National Science Foundation.

\clearpage

\section*{References}

\smallskip
\noindent\hspace{-0.5cm} 
Aaronson, M., Huchra, J.H., \& Mould, J.R.  1979, ApJ, 229, 1.

\smallskip
\noindent\hspace{-0.5cm}
Allen, R.J., \& Shu, F.H. 1979, ApJ, 227, 67.

\smallskip
\noindent\hspace{-0.5cm}
Barnes, J.E., \& Efstathiou, G.  1987, ApJ, 319, 575.

\smallskip
\noindent\hspace{-0.5cm}
Barnes, J.E., \& Hernquist, L.E.  1991, ApJ, 370, L65.

\smallskip
\noindent\hspace{-0.5cm} 
Binggeli, B., Sandage, A., \& Tammann, G.A.  1988, ARAA, 26, 509.

\smallskip
\noindent\hspace{-0.5cm} 
Binggeli, B., Sandage, A., \& Tarenghi, M.  1984, AJ, 89, 64.

\smallskip
\noindent\hspace{-0.5cm}
Blumenthal, G.R., Faber, S.M., Flores, R., \& Primack, J.R.  1986,
ApJ, 301, 27.

\smallskip
\noindent\hspace{-0.5cm}
Bothun, G.D., Impey, C.D., Malin, D.F., \& Mould, J.R.  1987, AJ, 94,
23.

\smallskip
\noindent\hspace{-0.5cm}
Carignan, C., \& Freeman, K.C.  1988, ApJ, 332, L33.

\smallskip
\noindent\hspace{-0.5cm}
Casertano, S., \& van Gorkom, J.H.  1991, AJ, 101, 1231.

\smallskip
\noindent\hspace{-0.5cm}
Cornell, M.E., Aaronson, M., Bothun, G.D., \& Mould, J.R.  1987, ApJS, 
64, 507.

\smallskip
\noindent\hspace{-0.5cm}
Courteau, S., de Jong, R.S., \& Broeils, A.H.  1996, ApJ, 457, L73.

\smallskip
\noindent\hspace{-0.5cm}
Davies, J.I., Disney, M.J., Phillipps, S., Boyle, B.J., \& Couch,
W.J.  1994, MNRAS, 269, 349.

\smallskip
\noindent\hspace{-0.5cm}
Davies, J.I., Phillipps, S., Cawson, M.G.M., Disney, M.J., \&
Kibblewhite, E.J.  1988$a$, MNRAS, 232, 239.

\smallskip
\noindent\hspace{-0.5cm}
Davies, J.I., Phillipps, S., \& Disney, M.J.  1988$b$, MNRAS, 231, 69P.

\smallskip
\noindent\hspace{-0.5cm}
Davis, M. \& Huchra, J.P.  1982, ApJ, 254, 437.

\smallskip
\noindent\hspace{-0.5cm}
de Blok, W.J.G., \& McGaugh, S.S.  1996, ApJ, accepted in letters.

\smallskip
\noindent\hspace{-0.5cm}
de Blok, W.J.G., \& McGaugh, S.S.  1997, MNRAS, submitted.

\smallskip
\noindent\hspace{-0.5cm}
de Blok, W.J.G., van der Hulst, J.M., \& Bothun, G.D. 1995, MNRAS,
274, 235.

\smallskip
\noindent\hspace{-0.5cm}
de Jong, R.S.  1996$a$, AAP, 313, 45.

\smallskip
\noindent\hspace{-0.5cm}
de Jong, R.S.  1996$b$, A\&AS, 118, 557.

\smallskip
\noindent\hspace{-0.5cm}
Dekel, A., \& Silk, J.  1986, ApJ, 303, 39.

\smallskip
\noindent\hspace{-0.5cm}
de Vaucouleurs, G.  1959, Hdb. d. Physik, 53, 311.

\smallskip
\noindent\hspace{-0.5cm}
Disney, M.J.  1976, Nature, 263, 573.

\smallskip
\noindent\hspace{-0.5cm}
Disney, M.J., \& Phillipps, S.  1983, MNRAS, 205, 1253.

\smallskip
\noindent\hspace{-0.5cm}
Dressler, A.  1980, ApJ, 236, 351.

\smallskip
\noindent\hspace{-0.5cm}
Driver, S.P., Phillipps, S., Davies, J.I., Morgan, I., \& Disney,
M.J.  1994, MNRAS, 268, 393.

\smallskip
\noindent\hspace{-0.5cm}
Efstathiou, G., \& Jones, B.J.T.  1979, MNRAS, 186, 133.

\smallskip
\noindent\hspace{-0.5cm}
Fall, S.M., \& Efstathiou, G.  1980, MNRAS, 193, 189.

\smallskip
\noindent\hspace{-0.5cm}
Ferguson, H.C., \& Sandage, A. 1988, AJ, 96, 1520.

\smallskip
\noindent\hspace{-0.5cm}
Fisher, J.R., \& Tully, R.B.  1975, AAP, 44, 151.

\smallskip
\noindent\hspace{-0.5cm}
Fisher, J.R., \& Tully, R.B.  1981, ApJs, 47, 139.

\smallskip
\noindent\hspace{-0.5cm}
Freeman, K.C.  1970, ApJ, 160, 811.

\smallskip
\noindent\hspace{-0.5cm}
Gunn, J.E.  1982, in Astrophysical Cosmology, ed. H.A. Br\"uck,
G.V. Coyne, and M.S. Longair (Vatican: Pontificia Academia
Scientiarum), p. 233.

\smallskip
\noindent\hspace{-0.5cm}
Hubble, E.  1936, ApJ, 84, 158.

\smallskip
\noindent\hspace{-0.5cm}
Impey, C.D., Bothun, G.D., \& Malin, D.  1988, ApJ, 330, 634.

\smallskip
\noindent\hspace{-0.5cm}
Impey, C.D., Sprayberry, D., Irwin, M.J., \& Bothun, G.D,  1996,
ApJS, 105, 209.

\smallskip
\noindent\hspace{-0.5cm}
Irwin, M.J., Davies, J.I., Disney, M.J., \& Phillipps, S.  1990,
MNRAS, 245, 289.

\smallskip
\noindent\hspace{-0.5cm}
Jura, M.  1980, ApJ, 238, 499.

\smallskip
\noindent\hspace{-0.5cm}
Kormendy, J.  1977, ApJ, 217, 406.

\smallskip
\noindent\hspace{-0.5cm}
Kormendy, J.  1985, ApJ, 295, 73.

\smallskip
\noindent\hspace{-0.5cm}
Loveday, J., Peterson, B.A., Efstathiou, G., \& Maddox, S.J. 1992,
ApJ, 390, 338.

\smallskip
\noindent\hspace{-0.5cm}
Mannheim, P.D.  1993, ApJ, 419, 150.

\smallskip
\noindent\hspace{-0.5cm}
Marzke, R.O., Geller, M.J., Huchra, J.P., \& Corwin, H.G. Jr.  1994,
AJ, 108, 437.

\smallskip
\noindent\hspace{-0.5cm}
McGaugh, S.S.  1996, MNRAS, 280, 337.

\smallskip
\noindent\hspace{-0.5cm}
McGaugh, S.S., \& Bothun, G.D.  1994, AJ, 107, 530.

\smallskip
\noindent\hspace{-0.5cm}
McGaugh, S.S., Bothun, G.D., \& Schombert, J.M.  1995, AJ, 110, 573.

\smallskip
\noindent\hspace{-0.5cm}
Mestel, L.  1963, MNRAS, 126, 553.

\smallskip
\noindent\hspace{-0.5cm}
Milgrom, M. 1983, ApJ, 270, 365.

\smallskip
\noindent\hspace{-0.5cm}
Navarro, J.F., Frenk, C.S., \& White, S.D.M.  1996, ApJ, 462, 563.

\smallskip
\noindent\hspace{-0.5cm}
Nilson, P.  1973, Uppsala General Catalogue of Galaxies,
Roy. Soc. Sci. Uppsala.

\smallskip
\noindent\hspace{-0.5cm}
Peletier, R.F., \& Willner, S.P.  1992, AJ, 103, 1761.

\smallskip
\noindent\hspace{-0.5cm}
Persic, M., \& Salucci, P.  1988, MNRAS, 234, 131.

\smallskip
\noindent\hspace{-0.5cm}
Persic, M., Salucci, P., \& Stel, F.  1996, MNRAS, 281, 27.

\smallskip
\noindent\hspace{-0.5cm}
Ryden, B.S.  1988, ApJ, 329, 589.

\smallskip
\noindent\hspace{-0.5cm}
Ryden, B.S., \& Gunn, J.E.  1987, ApJ, 318, 15.

\smallskip
\noindent\hspace{-0.5cm}
Sandage, A., Binggeli, B., \& Tammann, G.A.  1985, AJ, 90, 1759.

\smallskip
\noindent\hspace{-0.5cm}
Sanders, R.H.  1984, AAP, 136, L21.

\smallskip
\noindent\hspace{-0.5cm}
Schechter, P.L.  1976, ApJ, 203, 297.

\smallskip
\noindent\hspace{-0.5cm}
Schwartzenberg, J.M., Phillipps, S., Smith, R.M., Couch, W.J.,, \&
Boyle, B.J.  1995, MNRAS, 275, 121.

\smallskip
\noindent\hspace{-0.5cm}
Schwarz, U.J. 1985, AAP, 142, 273.

\smallskip
\noindent\hspace{-0.5cm}
Shaya, E.J., \& Tully, R.B. 1984, ApJ, 281, 56.

\smallskip
\noindent\hspace{-0.5cm}
Sprayberry, D., Bernstein, G.M., Impey, C.D., \& Bothun, G.D.  
1995, ApJ, 438, 72.

\smallskip
\noindent\hspace{-0.5cm}
Sprayberry, D, Impey, C.D., Bothun, G.D., \& Irwin, M.J.  1995, AJ,
109, 558.

\smallskip
\noindent\hspace{-0.5cm}
Tully, R.B.  1988, AJ, 96, 73.

\smallskip
\noindent\hspace{-0.5cm}
Tully, R.B., \& Fisher, J.R.  1977, AAP, 54, 661.

\smallskip
\noindent\hspace{-0.5cm}
Tully, R.B., Shaya, E.J., Pierce, M.J.,\& Peebles, P.J.E.  1997,
ApJs, in preparation.

\smallskip
\noindent\hspace{-0.5cm}
Tully, R.B., Verheijen, M.A.W., Pierce, M.J., Huang, J.S., \&
Wainscoat, R.J.  1996, AJ, 112, 2471. (Paper I)

\smallskip
\noindent\hspace{-0.5cm}
Valentijn, E.A.  1990, Nature, 346, 153. 

\smallskip
\noindent\hspace{-0.5cm}
van Albada, T.S., \& Sancisi, R.  1986, Phil. Trans. R. Soc. Lond. A,
320, 447.

\smallskip
\noindent\hspace{-0.5cm}
van den Bergh, S.  1959, Publ. David Dunlap Obs., II, No. 5.

\smallskip
\noindent\hspace{-0.5cm}
van den Bergh, S.  1966, AJ, 71, 922.

\smallskip
\noindent\hspace{-0.5cm}
van der Kruit, P.C.  1987, AAP, 173, 59.

\smallskip
\noindent\hspace{-0.5cm}
Verheijen, M.A.W.  1997, Ph.D. Thesis, University of Groningen, in
progress.

\smallskip
\noindent\hspace{-0.5cm}
Wirth, A. \& Gallagher, J.S. III  1984, ApJ, 282, 85.

\smallskip
\noindent\hspace{-0.5cm}
Zwaan, M.A., van der Hulst, J.M., de Blok, W.J.G., \& McGaugh, S.S.
1995, MNRAS, 273, L35.

\smallskip
\noindent\hspace{-0.5cm}
Zwicky, F.  1942, Phys. Rev., 61, 489.

\smallskip
\noindent\hspace{-0.5cm}
Zwicky, F.  1964, ApJ, 140, 1467.

\clearpage

\begin{table}[t]
\caption{Mean Central Surface Brightnesses}
\begin{center}
\begin{tabular}{ccccccccc}
Filter & $C_{HSB}$ & N & $<\mu_0>_{HSB}$ & rms & $C_{LSB}$ & N & $<\mu_0>_{LSB}$ & rms \\
\hline
\multicolumn{9}{l}{Complete sample} \\
$B$          & 0.23 & 39 & 20.57 & $\pm 0.57$ & 1.00 & 23 & 22.65 & $\pm 0.49$ \\
$R$          & 0.52 & 39 & 19.54 & $\pm 0.52$ & 1.00 & 23 & 21.72 & $\pm 0.48$ \\
$I$          & 0.61 & 39 & 18.94 & $\pm 0.54$ & 1.00 & 23 & 21.21 & $\pm 0.52$ \\
$K^{\prime}$ & 1.00 & 38 & 17.11 & $\pm 0.58$ & 1.00 & 22 & 19.61 & $\pm 0.67$ \\
\hline
\multicolumn{9}{l}{Complete subsample of isolated galaxies} \\
$B$          &      & 24 & 20.60 & $\pm 0.52$ &      & 14 & 22.77 & $\pm 0.46$ \\
$R$          &      & 24 & 19.50 & $\pm 0.55$ &      & 14 & 21.87 & $\pm 0.41$ \\
$I$          &      & 24 & 18.94 & $\pm 0.54$ &      & 14 & 21.36 & $\pm 0.46$ \\
$K^{\prime}$ &      & 23 & 17.15 & $\pm 0.40$ &      & 13 & 19.85 & $\pm 0.39$ \\
\hline
\end{tabular}
\end{center}
\end{table}

\begin{table}[b]
\caption{Revised exponential disk fits.}
\begin{center}
\begin{tabular}{lccrcll}
      &            & \multicolumn{2}{c}{Paper I} & & \multicolumn{2}{c}{Revised} \\
PGC   & Filter     &   $\mu_0$   &   r$_d$   & &   $\mu_0$   &   r$_d$   \\
\hline
35202 & K$^\prime$ &    19.31    &    19.6   & &     19.9    &    26    \\
36686 & K$^\prime$ &    17.04    &     7.1   & &     18.0    & \phantom{0}9 \\
37073 & K$^\prime$ &    17.09    &     7.9   & &     17.8    &    11    \\
37136 & K$^\prime$ &    16.90    &     9.3   & &     17.3    &    10    \\
37550 & B          &    21.91    &    13.1   & &     22.26   &    14.5  \\
      & R          &    20.78    &    11.7   & &     21.19   &    13.0  \\
      & I          &    20.25    &    11.2   & &     20.63   &    12.3  \\
      & K$^\prime$ &    18.19    &     7.9   & &     18.56   & \phantom{0}8.8 \\
37553 & B          &    21.06    &    16.6   & &     21.42   &    18.0  \\
      & R          &    20.22    &    17.5   & &     20.60   &    19.2  \\
      & I          &    19.88    &    20.0   & &     20.30   &    22.5  \\
      & K$^\prime$ &    17.87    &    14.6   & &     18.34   &    17.1  \\
37618 & K$^\prime$ &    16.04    &     8.9   & &     16.4    &    10    \\
37719 & K$^\prime$ &    14.70    &    11.0   & &     17.0    &    18    \\
38068 & K$^\prime$ &    16.72    &    30.2   & &     16.9    &    33    \\
38392 & K$^\prime$ &    16.15    &    19.7   & &     15.82   &    18.0  \\
38440 & K$^\prime$ &    15.63    &    23.9   & &     16.02   &    26.3  \\
39237 & K$^\prime$ &    16.51    &     7.2   & &     16.74   & \phantom{0}7.7 \\
38503 & K$^\prime$ &    17.11    &    12.1   & &     17.8    &    15    \\
38507 & K$^\prime$ &    18.26    &     5.7   & &     19.1    & \phantom{0}9   \\
40228 & K$^\prime$ &    15.91    &    19.7   & &     16.8    &    25    \\
\hline
\end{tabular}
\end{center}
\end{table}

\clearpage

\begin{figure}
\centerline{
\psfig{figure=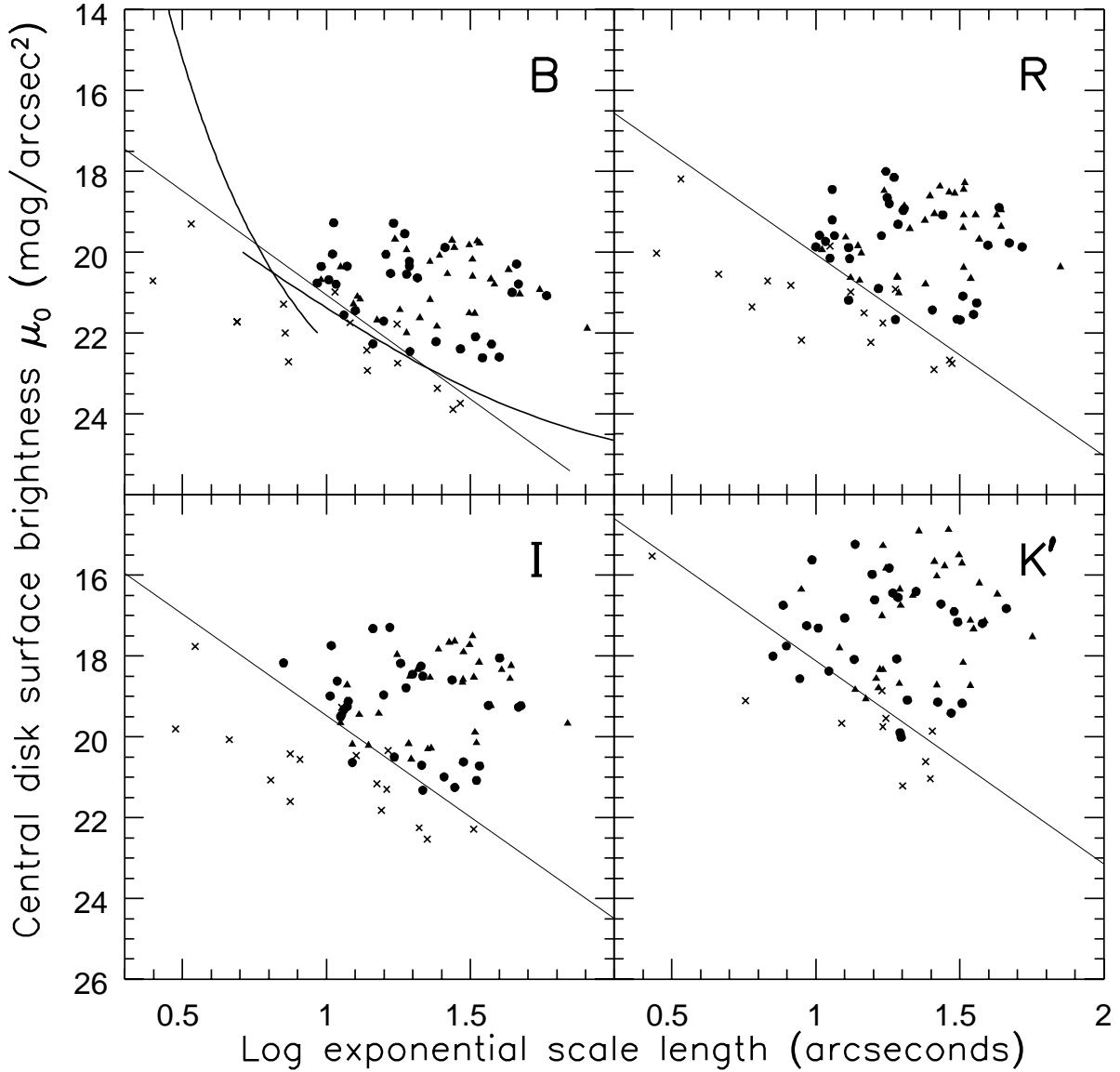,width=16cm}
}
\caption{
Central disk surface brightness $\mu_0$ versus exponential scale length
$r_d$.  Filled symbols: 62 galaxies of the complete sample with
$M_B^{b,i}<-16.5^m$; circles represent galaxies more face-on than
$60{\deg}$ and triangles represent galaxies more edge-on than this
limit.  Crosses: fainter galaxies.  Panels correspond to 
$B,R,I,K^{\prime}$ respectively.  Diagonal lines at $B=14.0^m$,
$R=13.1^m$, $I=12.5^m$, and $K^{\prime}=11.2^m$ 
represent the approximate limiting magnitudes for face-on systems.
For $B$, which is closest to the photographic band
used in the sample selection, two limiting visibility curves have been
superimposed.  Both curves are drawn on the assumption
that completion is limited by the surface brightness isophote
$\mu_{lim}^B=25.5$.  Galaxies with isophotal magnitudes brighter than
$14.5^m$ lie above the bottom curve and galaxies that are larger than
1~arcmin at the limiting isophote lie to the right of the steep curve
at the left of the figure.
} \label{1}
\end{figure}

\begin{figure}
\centerline{
\psfig{figure=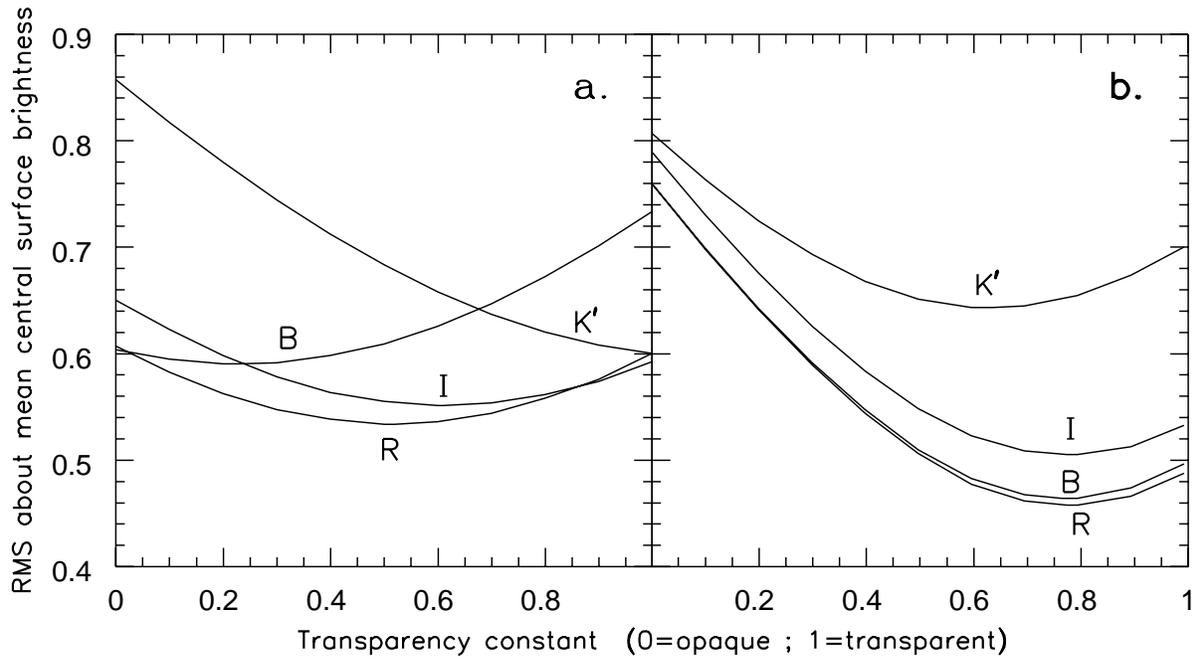,width=16cm}
}
\caption{
Dispersion about the mean central surface brightness as the
transparency constant $C$ is varied.  Separate curves are given for
the $B,R,I,K^{\prime}$ passbands.  Panel $a$ relates to the HSB
sub-sample and panel $b$ relates to the LSB sub-sample.
} \label{2}
\end{figure}

\begin{figure}
\centerline{
\psfig{figure=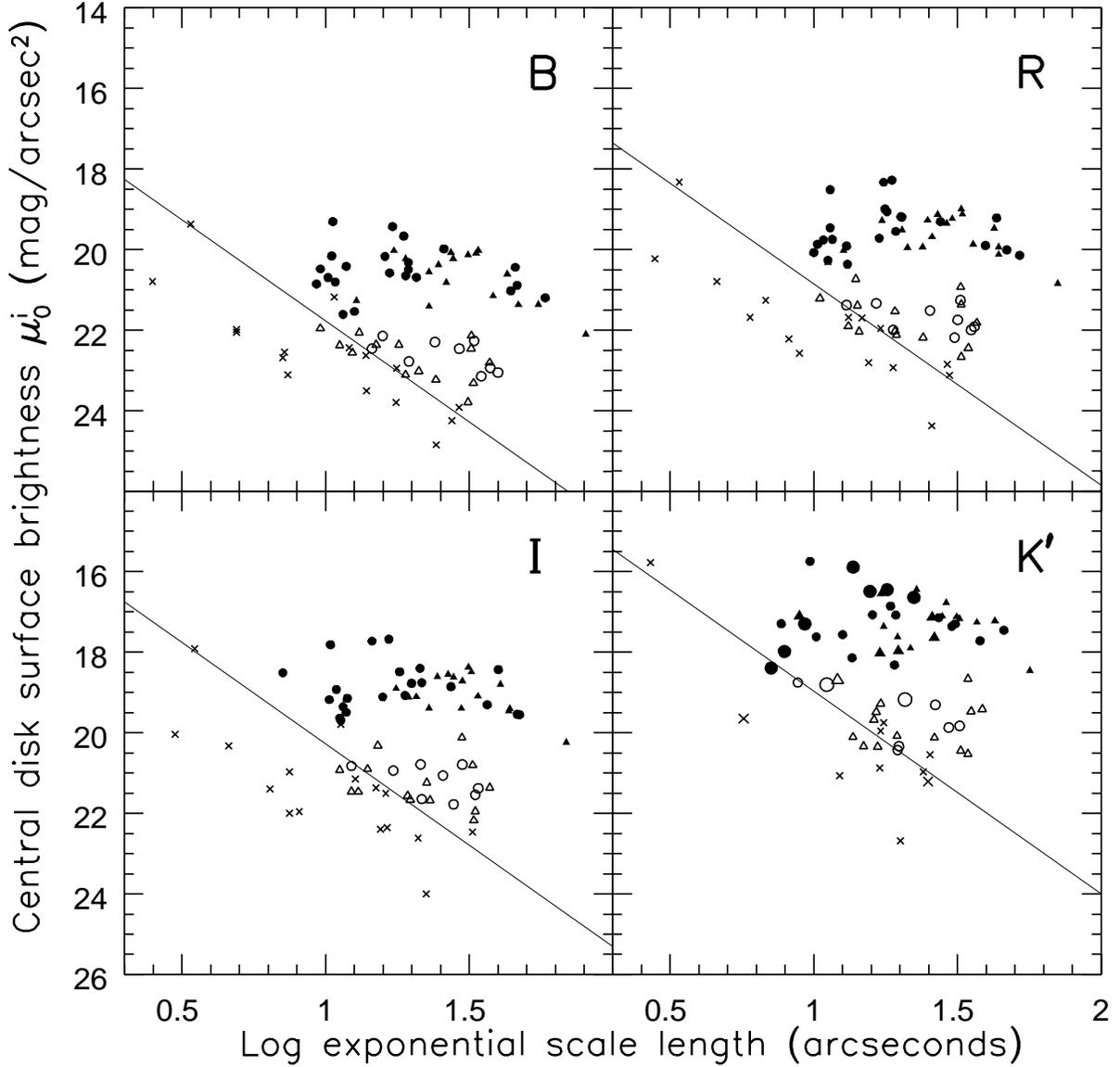,width=16cm}
}
\caption{
Inclination adjusted central disk surface brightness $\mu_0^i$ versus
exponential scale length $r_d$.  Now the complete sample is split between
HSB with $\mu_0^{K^{\prime},i} < 18.5$ (filled symbols), and LSB with
$\mu_0^{K^{\prime},i} > 18.5$ (open symbols).  Again, circles
represent galaxies more face-on than $60{\deg}$ and triangles
represent galaxies that are more edge-on than $60{\deg}$.  The panels
correspond 
to $B,R,I,K^{\prime}$.  Diagonal lines represent approximate limiting
magnitudes, accounting for mean inclination and color
transformations.  Larger symbols in the $K^{\prime}$ panel identify 
galaxies with bulges: concentration index $C_{82} > 5$.  
} \label{3}
\end{figure}

\begin{figure}
\centerline{
\psfig{figure=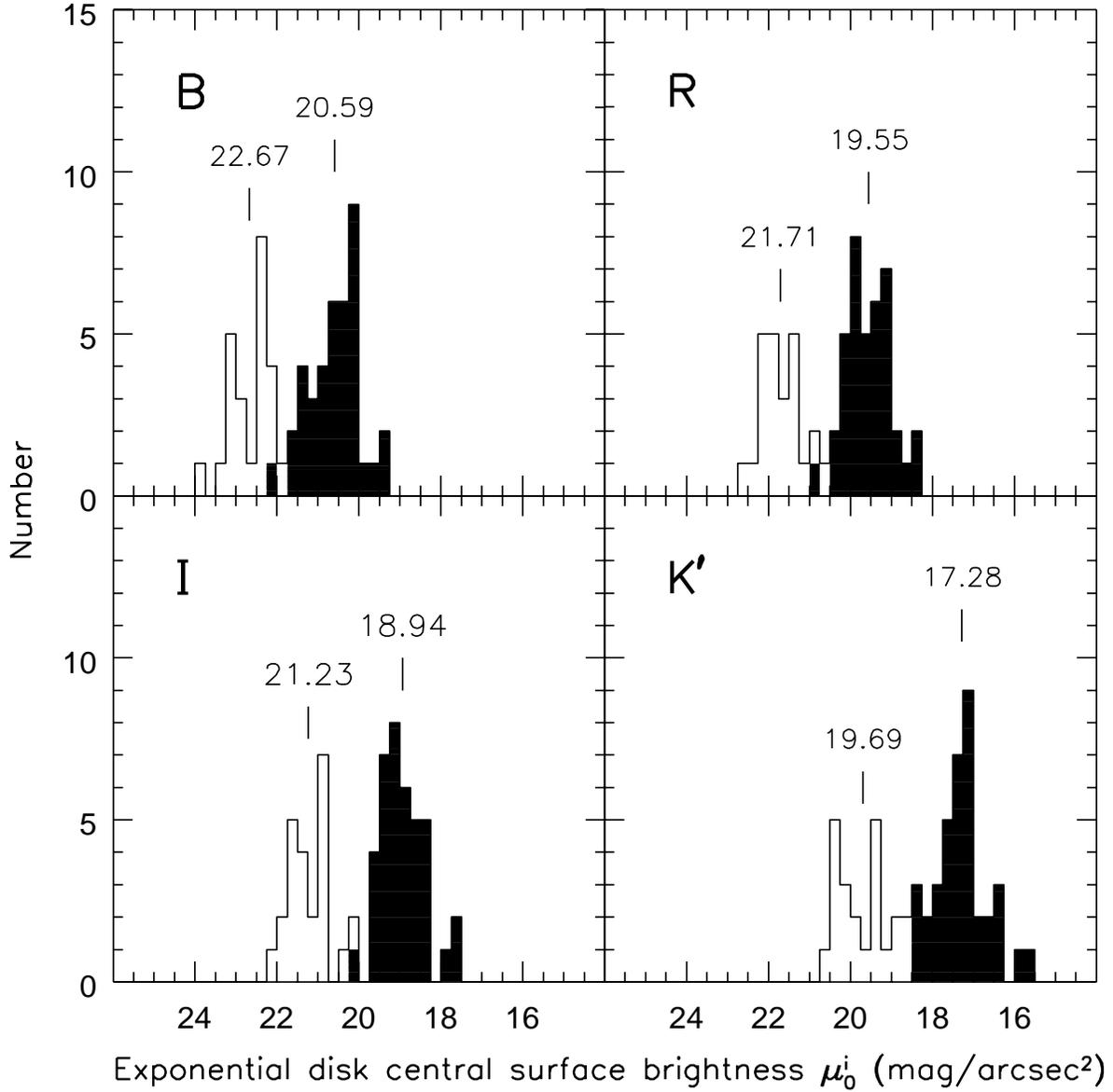,width=16cm}
}
\caption{
Histogram of inclination adjusted central disk surface brightnesses
for $B,R,I,K^{\prime}$.  The HSB component of the histogram is
filled.  Means for the HSB and LSB components are indicated.
NGC~3718 is identified as HSB at $K^{\prime}$ but lies with the LSB
at $B,R,I$.
} \label{4}
\end{figure}

\begin{figure}
\centerline{
\psfig{figure=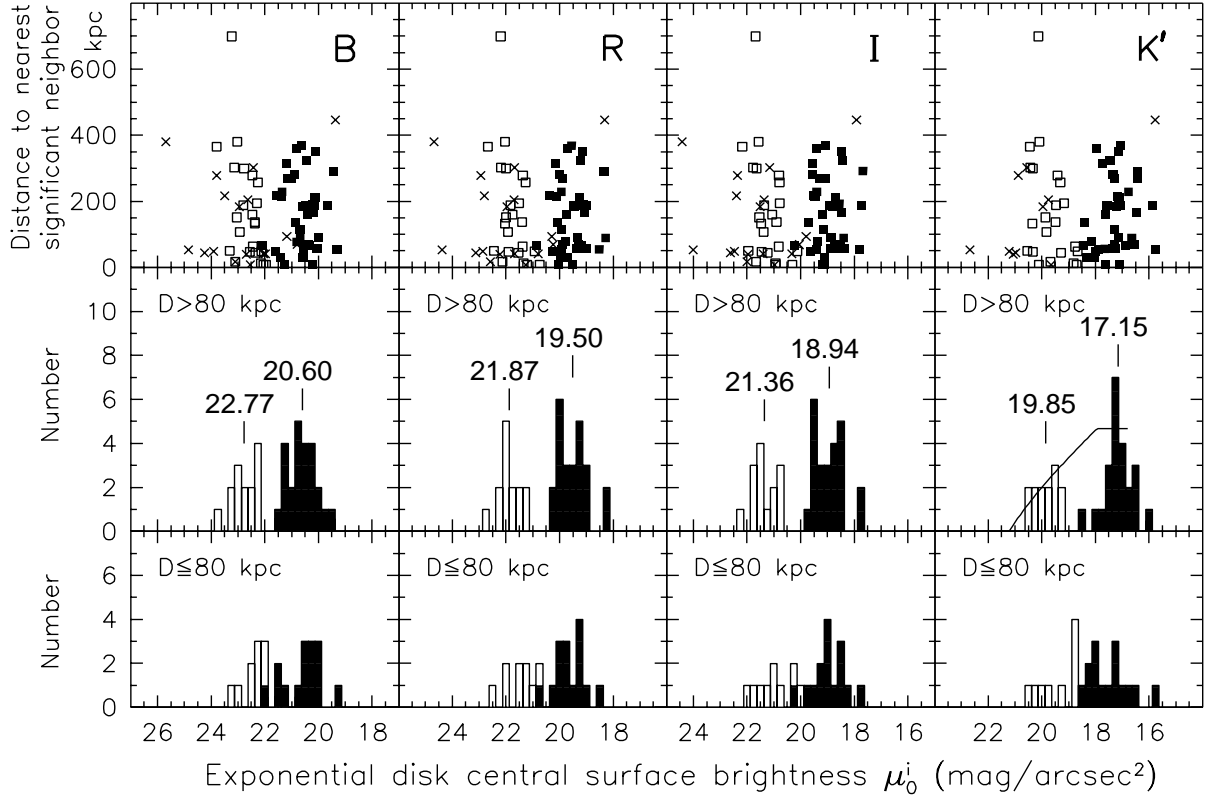,width=16cm}
}
\caption{
The effect of nearest neighbors on surface brightness.  In the top panel
for each of $B,R,I,K^{\prime}$, central disk surface brightnesses are 
plotted against the 
distance to the nearest neighbor with $L_{neighbor} > 0.1 L_{galaxy}$.
Closed symbols: HSB; open symbols: LSB; crosses: too faint for complete
sample.  The bottom two panels for each passband are decompositions of 
the histograms
of Fig.~4, complete sample only.  The middle panels gives the 
histogram for the subset with nearest projected neighbor farther than
80~kpc and the bottom panels gives the histogram for galaxies with a
projected neighbor closer than 80~kpc.  The curve in the 
$K^{\prime}$ middle panel
illustrates the completeness expectation if the surface
brightness--scale length plane is uniformly populated in the interval
$0.8<{\rm log} r_d<1.6$ and $\mu_0^{K^{\prime},i}>17^m$.  The
normalization of the maximum is given by the average of the three bins
at the peak of the HSB distribution.
} \label{5}
\end{figure}

\begin{figure}
\centerline{
\psfig{figure=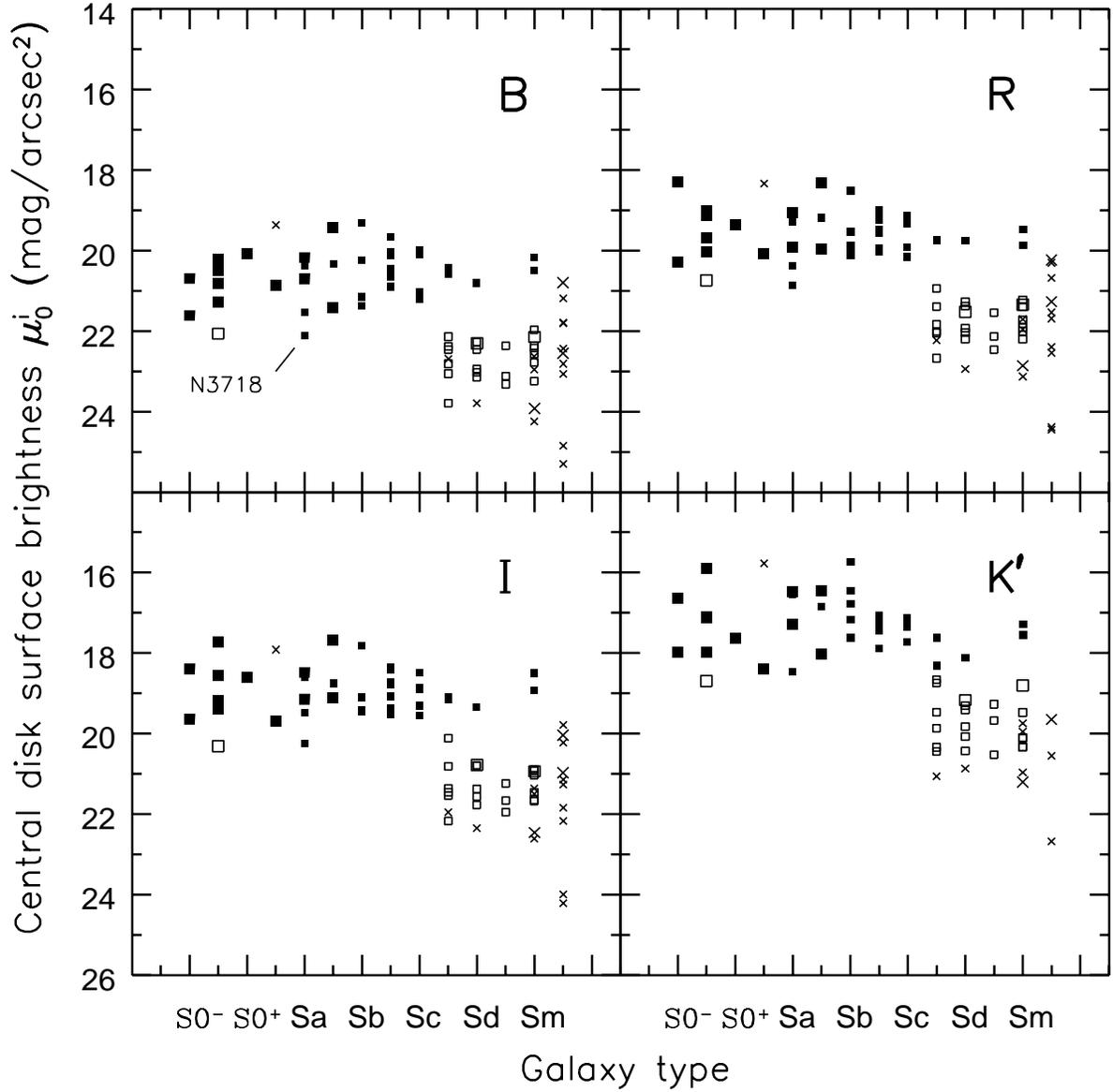,width=16cm}
}
\caption{
Correlation between surface brightness families and morphological
types.  Filled symbols: HSB; open symbols: LSB, crosses: fainter than 
complete sample limit.  Large symbols denote galaxies with bulges:
concentration index $C_{82} > 5$.
NGC~3718 is the unusual galaxy discussed in connection with the type~6, 
large low surface brightness systems.  The S0 galaxy identified as LSB
by the delineation between surface brightness regimes at 
$\mu_0^{K^{\prime},i} = 18.5$ is NGC~4117.
Otherwise, the LSB systems are 
typed S$cd$ or later.  The central disk surface brightnesses are similar
for bulge and non-bulge HSB systems.
} \label{6}
\end{figure}

\begin{figure}
\centerline{
\psfig{figure=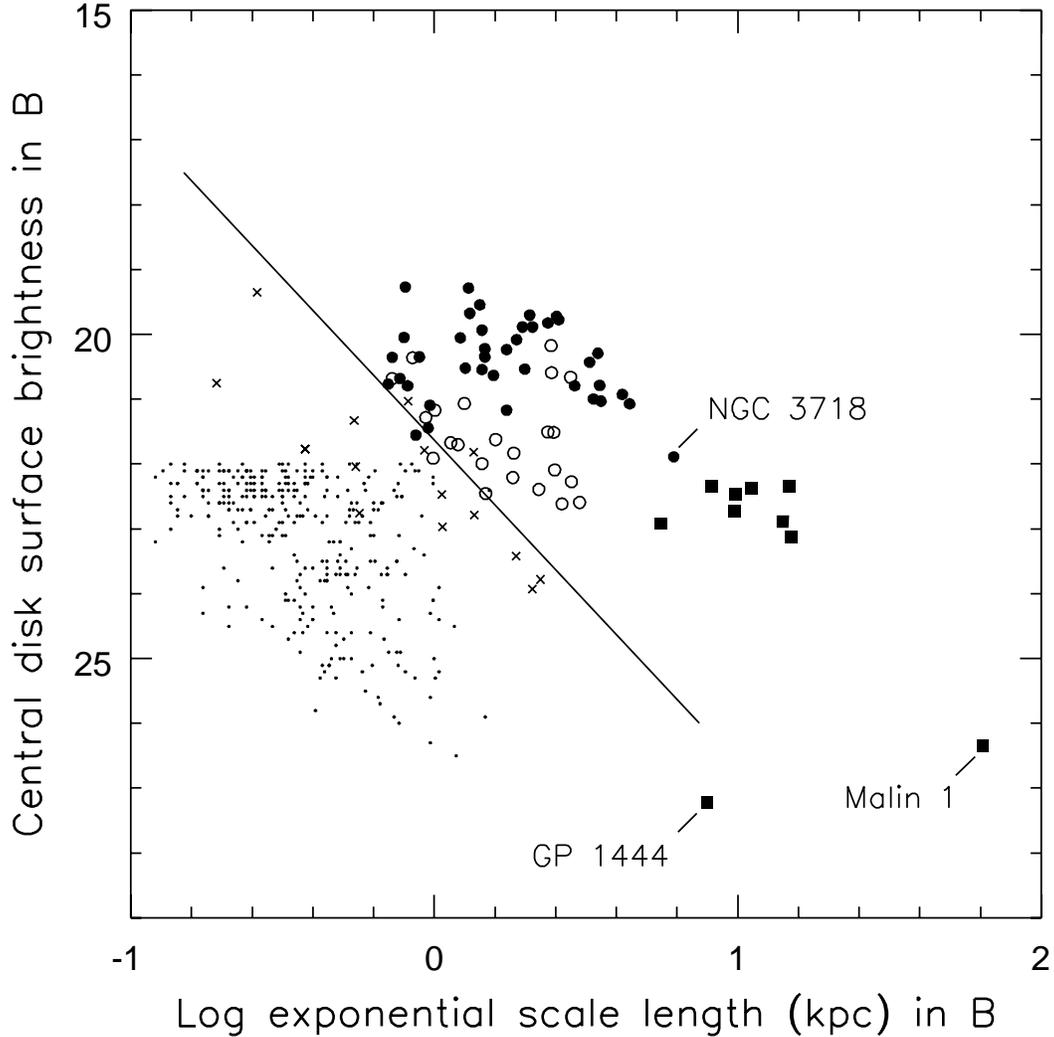,width=14cm}
}
\caption{
Comparison with $\mu_0-r_d$ information from the literature.  Data is
in $B$-band without inclination adjustments to make the broadest
comparison.  The  circles and straight line are the same information
shown in Fig.~1 ($B$ panel) except the HSB and LSB galaxies are now
filled and open symbols, respectively.  The small dots locate the low
surface brightness galaxies cataloged in the direction of the Fornax
Cluster by Davies et al. (1988$a$) and Irwin et al. (1990).  The points
are plotted assuming Fornax is 11\% more distant than Ursa Major.
There is confusion from background contamination for the shorter
scale length objects with $22 < \mu_0 \la 23$ but most of the fainter
surface brightness objects should be in the Fornax Cluster.  The large
filled squares identify the `giant' low surface brightness spiral
galaxies studied by Sprayberry et al. (1995) plus the two extreme systems
Malin~1 (Bothun et al. 1987) and GP~1444 (Davies et al. 1988$b$).  The
unusual galaxy NGC~3718 is identified in the figure.
} \label{7}
\end{figure}

\begin{figure}
\centerline{
\psfig{figure=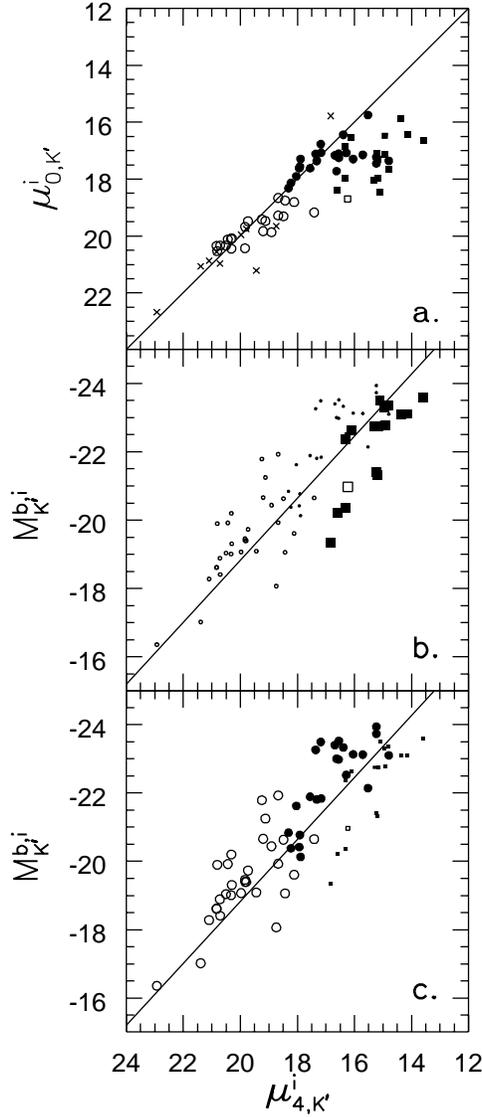,width=6.5cm}
}
\caption{
Correlations with the $K^{\prime}$ surface brightness within
$4^{\prime\prime}=300$~pc of the center of galaxies.  ($a$) Comparison
of the inclination corrected
exponential disk central surface brightness, $\mu_0^i$, and the
inclination corrected surface
brightness of the bulge plus disk in the central $4^{\prime\prime}$,
$\mu_4^i$. 
HSB: filled symbols; LSB: open symbols; $T \le 2$: boxes; $T \ge 3$:
circles; galaxies not in the complete 
sample: crosses.  The galaxies with substantial bulges 
($\mu_4^i \ll \mu_0^i$) almost all have types $\leq {\rm S}ab$.  ($b$)
Correlation between $\mu_4^i$ and absolute magnitude at $K^{\prime}$
with emphasis on early types.
Filled symbols: HSB; open symbols: LSB; big squares: type 
$\leq {\rm S}ab$; little circles: type $\geq {\rm S}b$.  The straight line is
the regression on {\it all} galaxies which minimizes the scatter in
$\mu_4^i$.  
The one open square is for NGC~4117, nominally LSB.  The only square to
the left of the regression line for the entire sample is NGC~3729, the
galaxy interacting with NGC~3718 and typed S$ab$.  
($c$) Same as panel~$b$ but with emphasis on late types.  Now circles
are big and squares are small.
} \label{8}
\end{figure}

\begin{figure}
\centerline{
\psfig{figure=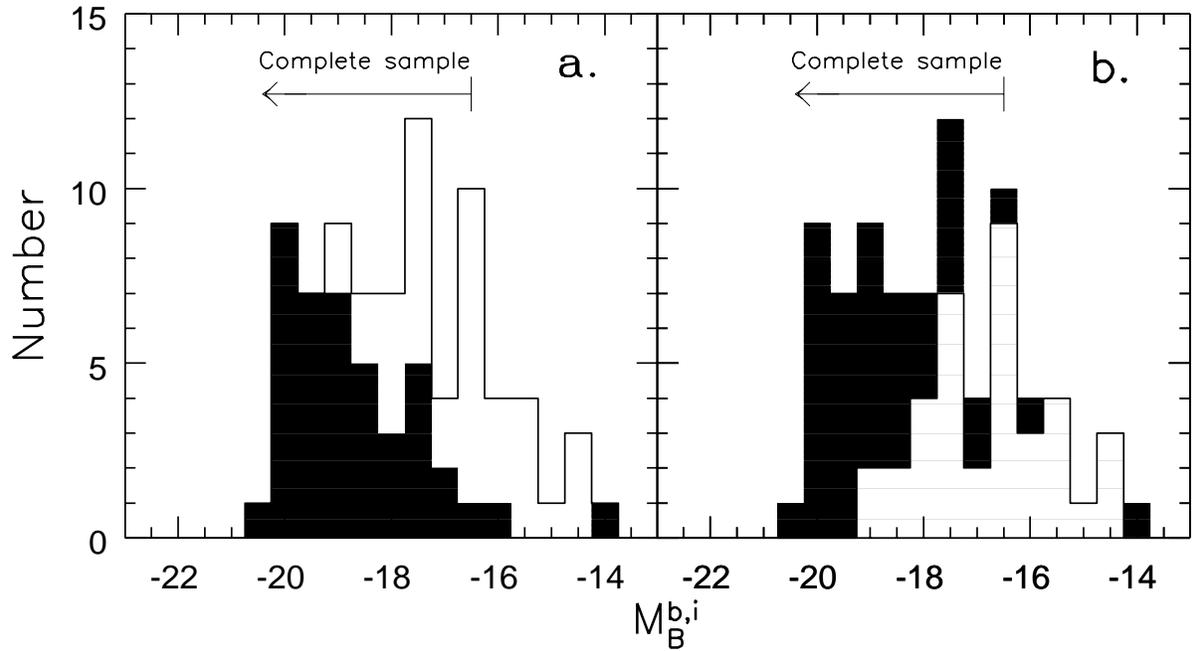,width=16cm}
}
\caption{
Luminosity functions in $B$-band.  ($a$) The HSB sub-sample is shaded
and embedded in the luminosity function for the entire sample.  There
is completion to $M_B^{b,i} = -16.5^m$.  The cutoff of the HSB
luminosity function by $M_B^{b,i} \sim -17^m$ is unlikely to be an
observational artifact.  ($b$) The LSB sub-sample is shown as an open
histogram embedded in the ensemble function.  The LSB component begins
at $M_B^{b,i} \sim -19^m$ and rises steeply to the completion limit.
} \label{9}
\end{figure}

\begin{figure}
\centerline{
\psfig{figure=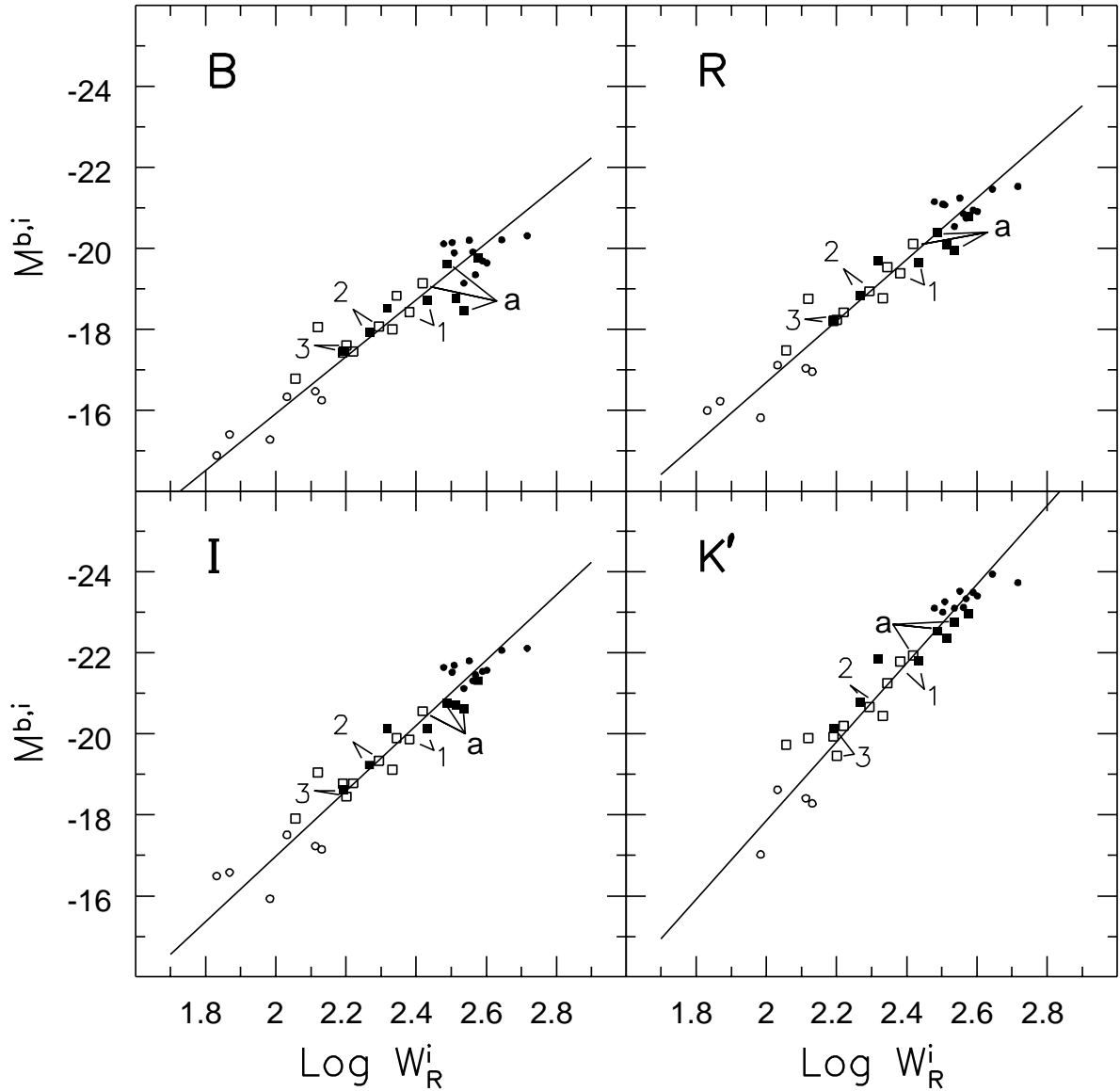,width=16cm}
}
\caption{
Luminosity--line width correlations are shown for 34 galaxies in 
the four passbands.  HSB: filled symbols; LSB: open symbols.
Galaxies in the overlap region $-23^m < M_{K^{\prime}}^{b,i} < -19^m$ 
($-19^m < M_B^{b,i} < -17^m$) are
indicated by box symbols.  Three pairs and a triplet given special 
attention are
noted by labels 1,2,3 and $a$.  The straight lines are
regressions which minimize scatter in line widths.
} \label{10}
\end{figure}

\begin{figure}
\centerline{
\psfig{figure=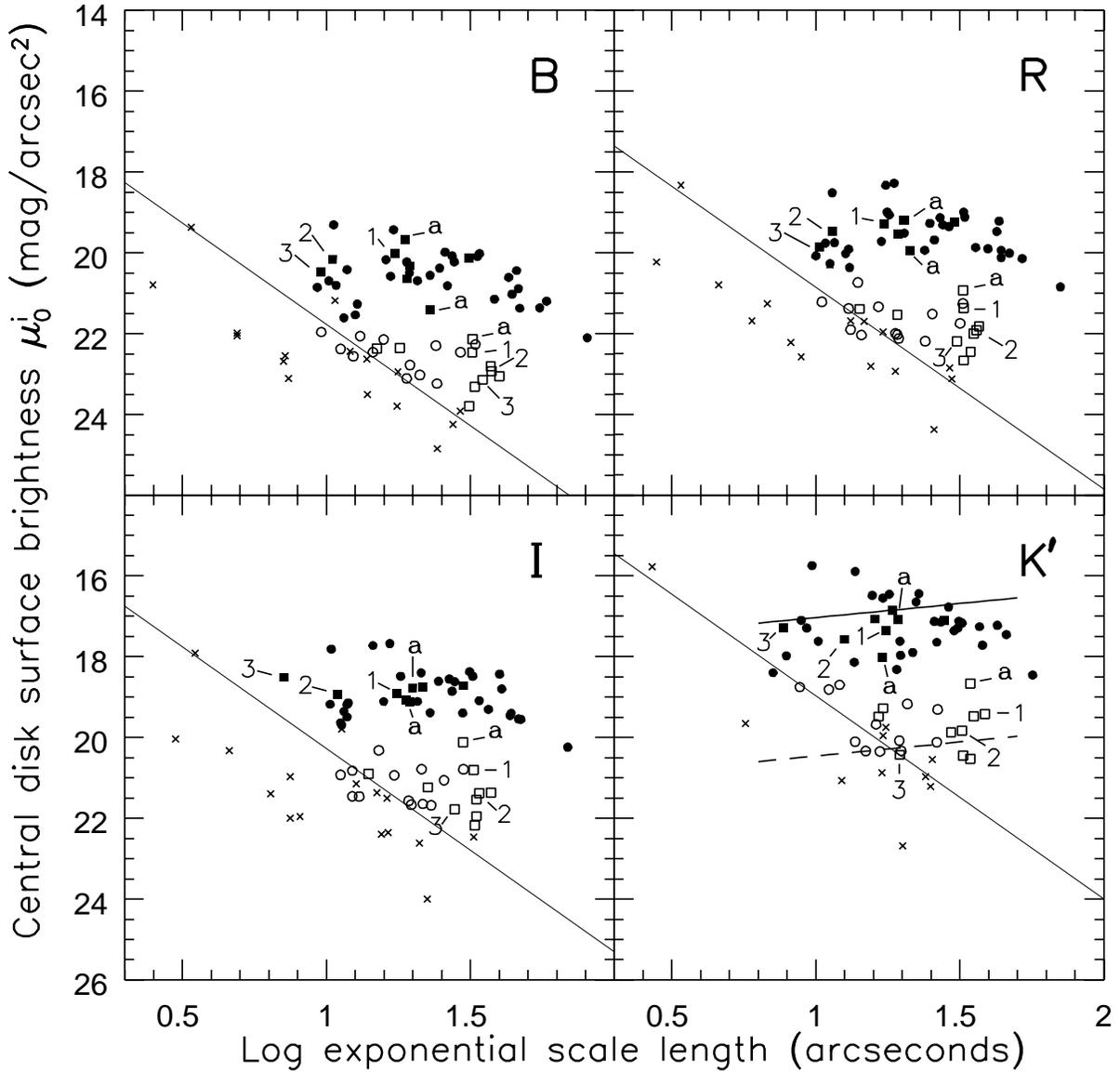,width=16cm}
}
\caption{
Repetition of Fig.~3 except the galaxies identified with box symbols
in Fig.~10 are identified with the same symbols here.  Small HSB 
galaxies and large LSB galaxies with similar luminosities also have
similar maximum rotation velocities.  Three pairs and a triplet given 
special attention are identified as in Fig.~10.  In the $K^{\prime}$
panel, the solid line is the locus of the relation 
$V_{max}^{disk}/0.5 W_R^i = 2/3$ while the dashed line correspondes to
$V_{max}^{disk}/0.5 W_R^i = 1/3$.
} \label{11}
\end{figure}

\begin{figure}
\centerline{
\psfig{figure=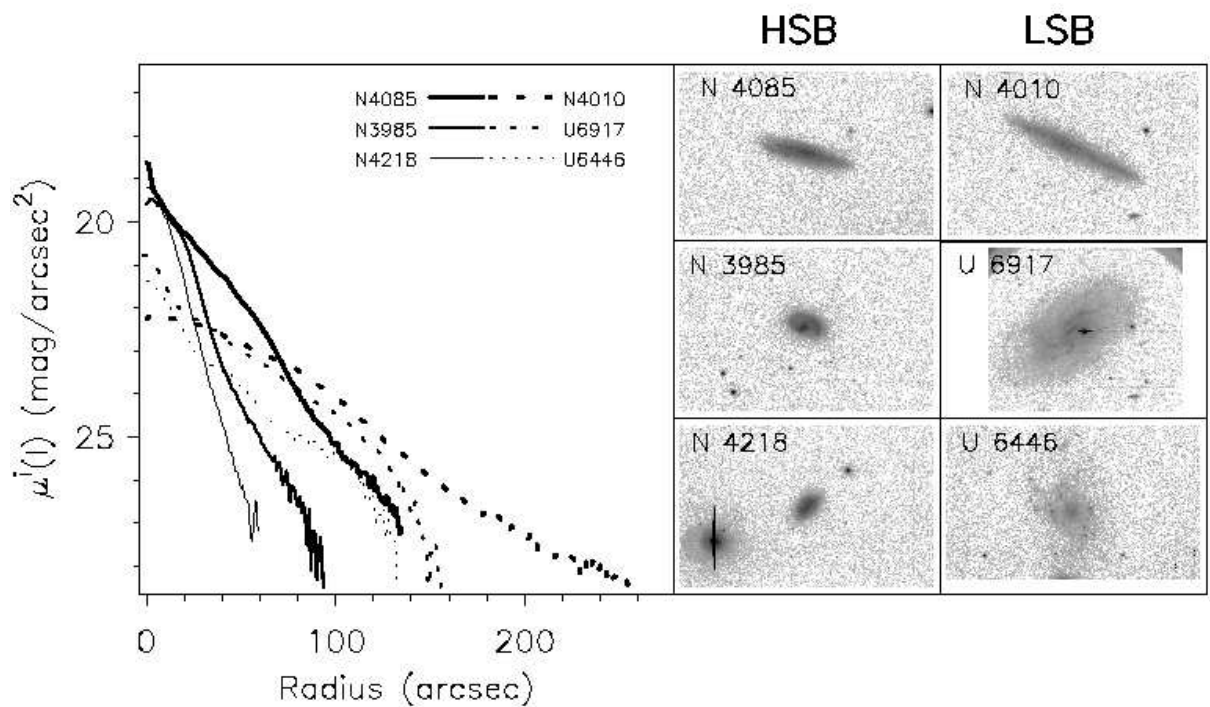,width=16cm}
}
\caption{
$I$-band inclination adjusted surface brightness versus radius for the three
HSB--LSB pairs of galaxies identified in Figs. 10 and 11.  The $B$
images of the pairs are shown in the inset at common scales. Pair~1:
NGC~4085/NGC~4010. Pair~2: NGC~3985/UGC~6917. Pair~3: NGC~4218/UGC~6446.
} \label{12}
\end{figure}

\begin{figure}
\centerline{
\psfig{figure=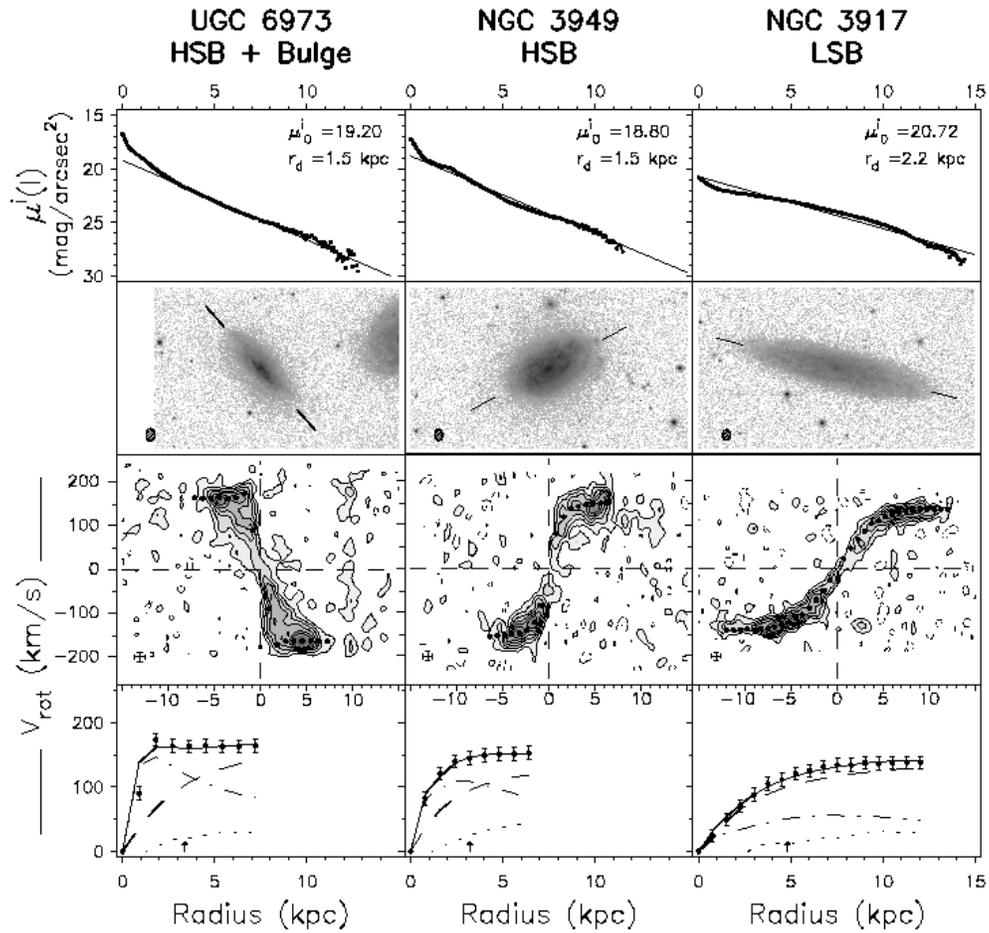,width=13cm}
}
\caption{
Rotation curve decompositions.  Examples of the three distinct classes of 
disk galaxies are presented in each of the vertical groups.  On the left
is UGC~6973, a system with a HSB exponential disk and a central bulge.  In the
middle is NGC~3949, a system with a HSB disk but no appreciable bulge.  On the
right is NGC~3917, a system with a LSB disk and no bulge.  In each case, the
horizontal axes are position in kpc.  In the top and bottom panels the origin
with respect to the nucleus is at the left axis, while in the middle panels the
origin with respect to the nucleus is at the center of the plots.  Surface
brightnesses at $I$, corrected for inclination, are shown in the top panels.
Images at $B$ are shown in the second row.  The major axes are indicated, as well
as the FWHM beam of the HI observations.  The velocity-position decomposition 
of the HI observations is seen in the panels of the third row.  Velocities
averaged over annuli are given as dots.  The rotation curve decompositions
are provided in the bottom panels.  The solid lines with error bars illustrate
the observed rotation curves.  Dot--dashed curves illustrate the amplitude of
rotation expected from the observed distributions of light and $M/L=0.4$ at
$K^{\prime}$.  The dotted curves illustrate the contribution expected from
the gas component.  The dashed curves demonstrate the residual contribution
attributed to a dark matter halo.  Isothermal spheres are used in this 
modelling.
} \label{13}
\end{figure}

\begin{figure}
\centerline{
\psfig{figure=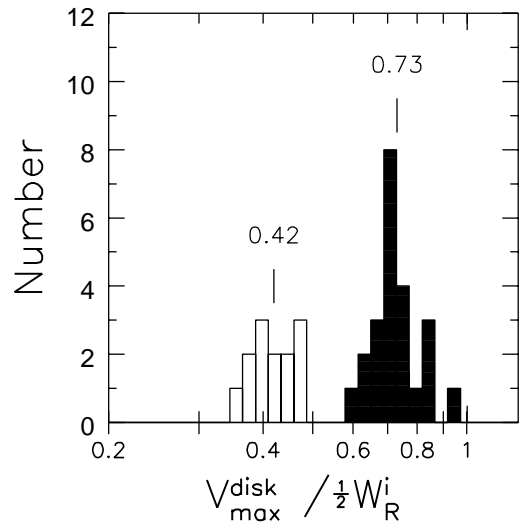,width=7cm}
}
\caption{
Histogram of the ratio $V_{max}^{disk}/0.5 W_R^i$ based on photometric 
properties.  Isolated sub-sample only (nearest important neighbor $>80$~kpc
in projection).  Values of $V_{max}^{disk}$ follow from the properties of
the exponential disk and $M/L_{K^{\prime}} = 0.4 M_{\odot}/L_{\odot}$.
The relation between $W_R^i$ and photometric parameters is given by the 
luminosity--line width correlation.  Filled histogram: HSB systems;
open histogram: LSB systems.  Mean values for each family are indicated.
} \label{14}
\end{figure}

\end{document}